\documentclass[manuscript]{emulateapj}
\usepackage{amsmath}     
\usepackage{wasysym}           
\usepackage{graphicx}
\usepackage{amssymb}
\usepackage{epstopdf}
\usepackage{mathrsfs}
\usepackage{natbib}
\begin{document}
\title{Hyperaccretion during tidal disruption events: weakly bound debris envelopes and jets}
\author{Eric R. Coughlin\altaffilmark{1} and Mitchell C. Begelman\altaffilmark{1}}
\affil{JILA, University of Colorado and National Institute of Standards and Technology, UCB 440, Boulder, CO 80309}
\email{eric.coughlin@colorado.edu, mitch@jila.colorado.edu}
\altaffiltext{1}{Department of Astrophysical and Planetary Sciences, University of Colorado, UCB 391, Boulder, CO 80309}

\begin{abstract}
After the destruction of the star during a tidal disruption event (TDE), the cataclysmic encounter between a star and the supermassive black hole (SMBH) of a galaxy, approximately half of the original stellar debris falls back onto the hole at a rate that can initially exceed the Eddington limit by orders of magnitude.  We argue that the angular momentum of this matter is too low to allow it to attain a disk-like configuration with accretion proceeding at a mildly super-Eddington rate, the excess energy being carried away by a combination of radiative losses and radially distributed winds. Instead, we propose that the infalling gas traps accretion energy until it inflates into a weakly-bound, quasi-spherical structure with gas extending nearly to the poles. We study the structure and evolution of such ``Zero-Bernoulli accretion" flows (ZEBRAs) as a model for the super-Eddington phase of TDEs. We argue that such flows cannot stop extremely super-Eddington accretion from occurring, and that once the envelope is maximally inflated, any excess accretion energy escapes through the poles in the form of powerful jets. {We compare the predictions of our model to \emph{Swift} J1644+57, the putative super-Eddington TDE, and show that it can qualitatively reproduce some of its observed features.} Similar models, including self-gravity, could be applicable to gamma-ray bursts from collapsars and the growth of supermassive black hole seeds inside quasi-stars. 
\end{abstract}


\section{Introduction}
Tidal disruption events (TDEs), encounters between a star and a massive ($\gtrsim 10^{5}M_{\astrosun}$) black hole in which the star passes within the tidal radius of the hole $r_t = R_*(M_h/M_*)^{1/3}$, where $R_*$ is the stellar radius, $M_*$ is its mass, and $M_h$ is the mass of the SMBH, have interested the astronomical community for decades.  Initial studies of TDEs focused on their potential for generating the luminosities observed in active galactic nuclei \citep{fra76,fra78}. While this pursuit fell by the wayside {(but see \citealt{mil06})}, TDEs continue to be useful for determining the presence of black holes within galactic centers \citep{lac82,ree88}. Many investigations, both computational and analytical, have been undertaken over the last forty years to elucidate the dynamics of the interaction between a star and a black hole and the luminosities associated with the resultant TDE \citep{car83,eva89,loe97,kim99,tch13,gui13}.

The earliest studies of the physics of the disruption noted that, due to the difference in the gravitational potential across the star, nearly half of the progenitor mass is ejected from the system on hyperbolic orbits \citep{lac82,car82,hil88}.  The other half remains bound to the black hole, with the orbits initially Keplerian (meaning that pressure forces have not yet altered particle trajectories). After a few revolutions of the innermost material, which occurs shortly after the disruption, hydrodynamical effects begin to modify the flow. These interactions result in the heating of the debris.

One heating agent arises from the pressure distribution of the tidally-disrupted star, which causes some of the orbits to possess a substantial inclination angle to the orbital plane of the center of mass of the star.  When the material on these orbits intersects the mid-plane of the disk, occurring roughly at periapsis, shock heating will simultaneously increase the internal energy of the gas and damp the inclination angle of the orbit. Because the innermost orbit has specific energy $\mathscr{E} \approx GM_hR_*/r_t^2 = GM_h/2R_i$, we see that $R_i \approx (R_*/2)(M_h/M_*)^{2/3}$ is its semi-major axis.  For a solar-mass star and a million-solar-mass hole, the advance of periapsis at this orbit can amount to degrees.  Because the innermost debris has the highest velocity within the disk, after one revolution it will impact the slower-moving, outer material, generating a shock. The shock heating further increases the thermal energy of the material and circularizes the orbits. Finally, material {continues to fall back at a rate that can greatly exceed the Eddington limit for some time} \citep{eva89,str11}. This accretion stage pumps a significant amount of energy into the flow.

Many authors have modeled the disks of bound material produced by TDEs (e.g., \citealt{can90,can92}) using the standard $\alpha$ parametrization of the viscosity and considering the disk to be radiatively efficient and geometrically thin \citep{sha73}. However, the processes outlined in the previous paragraph add a large amount of thermal energy to the debris, which consequently ``puffs up" the disk in the vertical direction. The super-Eddington accretion also means that radiative diffusion is not effective in cooling the disk.  We therefore believe that the thin disk model is incapable of describing the bulk properties of the flow during the super-Eddington phase.  

\citet{loe97} invoked the low specific angular momentum of the debris to enable them to describe the gas as roughly spherical; they then assumed that the black hole accretes at a rate to match its Eddington luminosity. To make this model self-consistent, they assumed that the spherical envelope, with an isentropic equation of state and a steep density profile $\rho \sim r^{-3}$, surrounded a rotating inner accretion flow, which would have to have a much flatter (and possible even inverted) density profile to obtain the required accretion rate.  The boundary between these regions was assumed to lie at roughly the tidal disruption radius, where the mean specific angular momentum of the debris is approximately Keplerian. While this model requires substantial angular momentum transport within the inner accretion flow, it offers no explanation as to why angular momentum should not be transferred to the outer envelope. If this happened, the boundary between the two zones would move to smaller radii, requiring the density profile of the inner flow to become even flatter (or more inverted), while the effect of rotation on the outer envelope would remain small. It is difficult to see how such a configuration could continue to regulate its accretion rate to remain at the Eddington limit.  On the contrary, it seems likely that the energy generation rate would become supercritical, with the envelope absorbing the excess energy until it became unbound.

In an alternate approach, \citet{str09} forced the material to conform to a ``slim disk," \citep{abr88} and supposed that an outflow carried away unbound debris.  However, a slim disk also necessitates that the rate at which matter reaches the black hole is only mildly super-Eddington, and therefore a fair amount of mass must be contained at large radii -- in this case, far outside the tidal radius. The assumption of nearly-Keplerian orbits, built into the slim disk model, then implies that the total angular momentum required to support the flow is larger than the angular momentum available.

The past twenty years have seen the emergence of direct observational evidence to support the existence of TDEs and their associated accretion disks \citep{pir88}.  \emph{ROSAT} discovered the first potential candidates for TDEs in the X-ray \citep{bad96,kom99}.  Despite the fairly small set of statistics, astronomers used the \emph{ROSAT} data to tentatively validate the rate of $10^{-5}$ events per galaxy per year \citep{don02}. \emph{Chandra}, \emph{GALEX}, and \emph{XMM-Newton} surveys have followed up on the events discovered by \emph{ROSAT}, {demonstrating that the luminosity-decay relation scales roughly as $t^{-5/3}$, as expected from early theoretical studies of TDEs} (\citealt{phi89}, {but see \citet{lod09} and \citet{gui13} for arguments against this scaling; also see section 4)}; they also found a few new potential candidates \citep{kom04,hal04,gez08}.  Most recently, a flurry of analyses has followed the discovery of the $\gamma$-ray, X-ray, and radio transient \emph{Swift} J164449.3+573451 (hereafter J1644+57), which is popularly believed to be the result of a TDE \citep{bur11,blo11,can11}. These studies, in particular the X-ray and radio observations, not only demonstrated the existence of {a roughly} $t^{-5/3}$ power-law decline of the undoubtedly super-Eddington luminosity, but also confirmed the novel association of a relativistic jet with the TDE \citep{zau11,tch13}.  

The combination of super-Eddington luminosity and a powerful jet suggests that accretion onto the black hole is not self-regulated, in contrast to previous models.  Here we adopt a different approach to modeling the super-Eddington accretion phase of the geometrically thick disks produced by TDEs. We assume that the structure of the flow is regulated by its ratio of angular momentum to mass, which is quite sub-Keplerian between the vicinity of the black hole and the photospheric radius.  Such a flow loses the ability to regulate its accretion luminosity, and absorbs energy liberated near the black hole until it becomes very weakly bound. Instead of blowing itself apart, however, we conjecture that these marginally bound envelopes can persist, with the excess accretion energy emerging as a jet through the narrow rotational funnel. We propose that such a model is consistent with the existence of a jet in \emph{Swift} J1644+57; it also may be relevant (with gas self-gravity included) to the formation of gamma-ray burst jets in collapsars and to the rapid growth of black holes inside quasi-stars.  

In section 2 we illustrate the model that describes the fallback disks in super-Eddington TDEs and promote reasons as to why this model is appropriate.  In section 3 we use the results of section 2 to analyze a disk whose parameters (mass, angular momentum, etc.) are those of a typical TDE and we show that the internal structure of the disk depends only on those bulk parameters.  In section 4 we consider the inner regions of the disk, where general relativity is important, discuss the properties of the jet, delineate the temporal evolution of the disk properties{, and compare our model directly to the case of \emph{Swift} J1644+57}. We conclude and review the results in section 5.

\section{Zero-Bernoulli accretion model}
The disk of stellar material created by a super-Eddington TDE should be thick in the sense that its scale height is some substantial fraction of its radial extent. \citet{nar94} were among the first to discover self-similar solutions for the vertically-averaged density, pressure, and angular momentum in which the velocity distribution was proportional to Keplerian. One interesting consequence of their models was that the Bernoulli parameter, given by $B = \Omega^2r^2/2-GM_h/{r}+H$, where $\Omega$ is the angular velocity and $H$ is the enthalpy, was shown to be greater than zero. Because the Bernoulli function is a measure of the specific energy of the gas, this result implies that any parcel of gas given an initial kick away from the SMBH would have energy at infinity.

The latter point motivated \citet{bla99} to describe a flow consisting of an advection dominated accretion flow (with $\dot{M}$ a function of radius) and a pressure-driven wind, calling these states ``adiabatic inflow-outflow solutions" (ADIOS). The inflowing gas maintains a negative Bernoulli parameter by transferring mass, angular momentum, and energy to the wind. The model, however, requires some unspecified mechanism -- presumably some dissipative process -- to drive the outflow.  Moreover, the inflow zone shares the characteristic of slim disks that the specific angular momentum must be very close to Keplerian -- more than 87 percent of the Keplerian value for the case of $\gamma = 4/3$. To avoid a highly super-Eddington accretion luminosity, this angular momentum distribution must extend to radii far beyond the tidal disruption radius{. The} gas returning to the vicinity of the black hole following a TDE{, however,} has too little angular momentum to permit this.

Thus, the gas distribution during the super-Eddington phase of a TDE is likely to resemble the quasi-spherical envelope of the \citet{loe97} model, but without the ability to regulate its accretion luminosity to a value close to the Eddington limit.  The shock heating of different parts of the disk and the energy input of the accreting black hole both raise the internal energy of the material, with turbulence, convection and internal shocks distributing that energy fairly evenly throughout the disk.  Eventually there will come a point in time where the Bernoulli parameter approaches zero, leaving a marginally-bound, highly-inflated envelope.  

Any further augmentation of the energy would start to unbind material. The question is whether this unbound material is launched from a wide range of radii or from close to the black hole, where the energy is injected. In the case of an ADIOS, the large angular momentum contained in the flow allows the system to maintain a disklike geometry, with a large ``free" surface along which a wind can develop. But in the present case, for which $B$ approaches zero, the disk closes up to a vanishingly narrow funnel, leaving the outer, quasi-spherical surface as the only plausible location for the development of a wide-angle wind. 

However, the injection of energy from the accreting black hole occurs deep in the interior of the envelope, where timescales are much shorter than those throughout the bulk of the flow. The accretion energy, pumped into the gas at a rate that is highly supercritical, is thus unlikely to be able to be efficiently advected to the outer regions where a wind could regulate the super-Eddington luminosity. The only viable exhaust route left for the excess energy is then along the poles, where the surface of the inflated envelope closes. We thus propose that, at this point in the evolution of the fallback disk, where the accretion luminosity augments the binding energy of the envelope to the point where a wind would develop if there were a free surface, a jet carries away the excess energy.

In the situations we are considering, the mass of the black hole dominates the total mass of the system.  We can therefore approximate the gravitational potential by $\phi = -GM_h/r$, where $M_h$ is the mass of the SMBH and $r$ is the radial distance from the hole (we are neglecting any contribution from post-Newtonian gravity; see section 4.1 for a discussion of relativistic effects). In spherical coordinates with this potential and the Bernoulli parameter equal to zero, the momentum equations and the Bernoulli equation are, respectively,

\begin{equation}
\frac{1}{\rho}\frac{\partial{p}}{\partial{r}} = -\frac{GM_h}{r^2}+\frac{\ell^2\csc^2\theta}{r^3}, \label{rmom}
\end{equation}
\begin{equation}
\frac{1}{\rho}\frac{\partial{p}}{\partial\theta} = \frac{\ell^2\cot\theta\csc^2\theta}{r^2}, \label{thmom}
\end{equation}
\begin{equation}
-\frac{GM_h}{r}+\frac{\ell^2\csc^2\theta}{2r^2}+\frac{\gamma}{\gamma-1}\frac{p}{\rho} = 0, \label{bern}
\end{equation}
where $\ell$ is the specific angular momentum of the gas. In the final line we used a specific form for the enthalpy and assumed that the azimuthal velocity is much greater than the poloidal or radial velocity. Here $\gamma$ is the adiabatic index of the gas, generally between 4/3 and 5/3 depending on the relative contributions from radiation pressure and gas pressure. For most of what follows we will assume that $\gamma \approx 4/3$, as radiation pressure dominates the support of TDE debris against gravity during the super-Eddington phase. This fluid description is appropriate to a ZEro-BeRnoulli Accretion (ZEBRA) flow.

\subsection{Gyrentropic flow}
In \citet{bla04}, the authors described ADIOS disks as marginally stable to the H\o{i}land criteria. This assumption, verified numerically (Stone et al. 1999), demanded that the surfaces of constant Bernoulli parameter, angular momentum, and entropy all coincide; these surfaces are termed \emph{gyrentropes}.  While the H{\o}iland criteria determine a disks's stability to convection in the absence of magnetic fields, even a vanishingly small poloidal field can completely destabilize a differentially rotating disk that is stable to those criteria \citep{bal91,bal92,sto94}. We will now show, however, that the zero-Bernoulli assumption ensures the gyrentropicity of the flow, even in the presence of the magnetorotational instability (MRI).

One can show that $\nabla{B} = H\nabla\ln{S}+\Omega\nabla\ell$, where $H$ is the enthalpy, $S$ is the entropy, and $\Omega$ is the angular velocity of the gas \citep{bla04}. Thus, since $B \approx 0$, $\nabla{S} \propto -\nabla{\ell}$. This relationship implies that surfaces of constant $S$ are also those of constant $\ell$, which must then be surfaces of constant $B$. This type of disk is therefore also gyrentropic, the constancy of $B$ being the only assumption which led to that conclusion. Thus, while MRI may invalidate the assumption of gyrentropicity on the grounds of the H{\o}iland criteria, a constant-Bernoulli disk retains gyrentropic flow (assuming that the magnetic energy density is not large enough to substantially alter the dynamical equilibrium).

\subsection{Self-similar solutions}
From an analysis of equations \eqref{rmom}, \eqref{thmom}, and \eqref{bern}, one can show that the general solution of $\ell(r,\theta)$ could have any functional form that depends on $r$ and $\theta$ only through the combination $r\sin^2\theta$ (see Appendix A, notably equation \eqref{angmom}). When the envelope subtends a large range in radii, however, we expect the solution to have a roughly self-similar structure between the inner and outer boundaries of the disk.  \citet{bla04} derive the gyrentropic solutions for arbitrary Bernoulli parameter $B(\theta)/r$; the ZEBRA solutions are the special case with $B=0$. We will simply quote their findings here, and adapt our notation to one which is consistent with theirs.  For the density, pressure, and specific angular momentum (squared), respectively, we find

\begin{equation}
\rho(r,\theta) = \rho_0\bigg{(}\frac{r}{r_0}\bigg{)}^{-q}(\sin^2\theta)^{\alpha},
\end{equation}
\begin{equation}
p(r,\theta) = \beta\frac{GM_h\rho_0}{r}\bigg{(}\frac{r}{r_0}\bigg{)}^{-q}(\sin^2\theta)^{\alpha},
\end{equation}
\begin{equation}
\ell^{\,2}(r,\theta) = aGM_hr\sin^2\theta,
\end{equation}
where 

\begin{equation}
q \equiv 3/2-n ,
\end{equation}
\begin{equation}
\alpha = \frac{1-q(\gamma-1)}{\gamma-1}, \label{alphaeq}
\end{equation}
\begin{equation}
\beta = \frac{\gamma-1}{1\mathbf{+ \gamma}-q(\gamma-1)}, \label{betaeq}
\end{equation}
\begin{equation}
a = 2\frac{1-q(\gamma-1)}{1+\gamma-q(\gamma-1)}, \label{aeq}
\end{equation}
$r_0$ is some characteristic inner radius and $\rho_0$ is the density at that radius (and at the disk midplane).  The parameter $n$ is defined by Blandford and Begelman so that the accretion rate is proportional to $r^{n}$; mass-conserving accretion has $n=0$.

One interesting aspect of these solutions is that $n$, and therefore $q$, which describes how steeply the density and pressure fall off as functions of $r$, is not specified a priori, which introduces another degree of freedom into the models. In general, however, we require that the exponent of $\sin^2\theta$ remain positive, ensuring that the density and pressure do not go to infinity at the poles. We also expect that the energy produced in the disk should be a decreasing function of radius.  From the energy equation, we know that the luminosity is given by $L \sim \dot{M}v_r^2$, and assuming that the power is produced by gas in regions with velocity appropriate to that for free-fall, we find that $L \propto r^{n-1}$. These two restrictions then impose that $3/2-1/(\gamma-1) < n < 1$, which translates to $1/2 < q < 1/(\gamma-1)$. We see that, since the exponent of $\sin^2\theta$ is always greater than or equal to zero in our self-similar expressions, the density goes to zero only exactly at the poles. These solutions thus represent quasi-spherical envelopes. The angular momentum distribution is modified from that of Keplerian by the factor $a$, which is always less than unity for permissible values of $n$.

\citet{bla04} noted the additional degree of freedom contained in their solutions. They then went on to describe the physical scenarios appropriate to different values of $n$. In particular, different $n$ give rise to larger or lesser amounts of outflow, accretion rates, energy generation rates, etc. For our present considerations, however, a wind is unnecessary.  The question of the value of $n$ therefore merits some careful consideration.  In the next section we will see how the properties of the disrupted star and the black hole in a TDE determine this as-yet-undetermined parameter in our analysis. Interestingly, the ZEBRA models admit a wider range of $n$-values than the range ($0 < n < 1$) consistent with ADIOS models.

\section{ZEBRA models of TDE debris disks}
The structure and evolution of a ZEBRA model for a TDE are governed by the total mass and angular momentum of the envelope, which change as matter falls back and is accreted or expelled in a jet. The total angular momentum and mass of the fallback disk are, respectively, $\mathscr{L} = \int\ell\,\rho\,{dV}$ and $\mathscr{M} = \int\rho\,{dV}$, where $dV$ is an infinitesimal volume element and the integral is taken over the whole fluid. Using the formalism and notation of the previous section, these can be written

\begin{equation}
\mathscr{L} = \frac{4\pi\rho_0\sqrt{aGM_h}}{r_0^{\,-q}}\int_0^{\pi/2}\int_{r_0}^{\mathscr{R}}r^{-q+5/2}(\sin^2\theta)^{\alpha+1}\,{dr}\,{d\theta}, \label{mathL}
\end{equation}
\begin{equation}
\mathscr{M} = \frac{4\pi\rho_0}{r_0^{\,-q}}\int_0^{\pi/2}\int_{r_0}^{\mathscr{R}}r^{-q+2}(\sin^2\theta)^{\alpha+1/2}\,{dr}\,{d\theta}, \label{mathM}
\end{equation}
where $r_0$ is simultaneously the radius at which we specify the density and the inner radius of the disk, and $\mathscr{R}$ denotes the outer radius. Due to the influence of the black hole, we expect $r_0$ to be on the order of the Schwarzschild radius (or, more precisely, the location of the innermost stable circular orbit \citep{bar72}), and so we will write $r_0 = \chi\,2GM_h/c^2$, with $\chi$ a pure number of order a few. To determine the outer radius, we compare the ability of the disk to transport energy via advection to its ability to transport energy via radiative diffusion.

Although the photosphere of the envelope may be radiating at close to the Eddington limit, the amount of energy generated in the interior of the disk will generally be much greater than that able to be carried via diffusion; specifically, the luminosity carried into the polar regions exceeds the Eddington limit by a factor of order $\ln(\mathscr{R}/r_0)$ \citep{jar80,pac80,sik81}. The dominant mode of energy transport will therefore be turbulent advection. The advective flux can be written $F_a = ypv$, where $p$ is the pressure, $v$ is the local sound speed, and $y$ is a number less than or of order one that describes the efficiency of advection (since we are really concerned with the flux of enthalpy, which is $4p$ for a radiation-dominated gas, $y$ could conceivably be greater than 1). Since $p \sim \rho{v^2}$ and the advective luminosity is $L_a \sim 4\pi{r^2}F_a$, we have that $L_a \sim 4\pi{y}r^2\rho{v^3}$. When the saturated advective luminosity becomes roughly equal to the Eddington limit, radiative diffusion will become the dominant mode of energy transport, allowing the disk to cool and become thin. Symbolically we have $4\pi{y_{max}}r^2\rho{v^3} \sim 4\pi{G}cM_h/\kappa$, where $\kappa$ is the relevant opacity. In this case we will use the opacity for electron scattering, given by $\kappa \approx 0.34$ cm$^2$/g for cosmological abundances. This definition is equivalent to that which defines the trapping radius -- the point in the flow at which the diffusion timescale equals the advective timescale \citep{beg78}. Fluid interior to this radius entrains photons, rendering them incapable of escaping.

In addition to having a magnitude, the advective flux has a directionality. Writing $\bold{F}_a = F_a\,\hat{n}$, the advective luminosity is obtained by integrating the dot-product of this vector over an area. Because we are concerned with the energy escaping from the hole, the relevant area is the two-sphere, and hence the only component of the flux relevant to the luminosity is that in the $\hat{r}$-direction. The quantity $\hat{n}\cdot\hat{r}$ will, in general, depend on $\theta$, and in fact we expect it to be less than one as much of the flux is transported into the polar regions. Because we are unaware of the specifics of the directional dependence of the flux, we will simply incorporate those uncertainties into our efficiency factor $y$, letting $\int{y\,p\,v\,\hat{n}\cdot\hat{r}dS} \equiv \bar{y}\int{p\,v\,dS}$, where $S$ is the two-sphere and $\bar{y}$ is an effective efficiency. Performing the integrations, we find that the outer radius is given by

\begin{equation}
\mathscr{R}^{-q+1/2} = \frac{2c}{\kappa\sqrt{\pi}}\frac{\Gamma(\alpha+3/2)}{\Gamma(\alpha+1)}\frac{r_0^{-q}}{\rho_0y\beta\sqrt{aGM_h}}. \label{goodR}
\end{equation}
The $\Gamma$-functions resulted from our integration of the angular dependence of $p$ over the two-sphere, and for simplicity we replaced $\bar{y}$ with $y$. Numerically we find that $1 < (2/\sqrt\pi)(\Gamma(\alpha+3/2)/\Gamma(\alpha+1)) < 2$ over permissible values of $\alpha$, so that its inclusion in our expression does not significantly alter our results. We will include the $\Gamma$-functions here, however, because they will simplify (visually) some of the relationships we will describe in later sections. To offer some insight into the meaning of the previous expression, note that it may be written 

\begin{equation}
\mathscr{R} \simeq (\frac{v_0}{c/\tau_0})^{\frac{1}{q-1/2}}r_0,
\end{equation}
where $\tau_0$ is the optical depth and $v_0$ is the local Keplerian velocity, both evaluated at $r_0$.  From this form of the equation, it is evident that photons must be trapped at $r_0$, namely the inequality $v_0 > c/\tau_0$ must hold, to ensure that our assumption about the radiative inefficiency of the flow be upheld.

Recall that $1/2 < q < 1/(\gamma-1)$, a restriction that resulted from requiring the density to be finite at all angles and the energy generation rate to increase inwards.  The adiabatic index of the gas will generally be between 4/3, and 5/3, meaning that $1/(\gamma-1) < 3$, and consequently $1/2 < q < 3$. Returning to equations \eqref{mathL} and \eqref{mathM}, we see that this range of $q$ will always leave the lower bound on the radial integration, namely $r_0$, relatively unimportant (unless $q$ is exactly 3, a case that we will have to consider separately) if $\mathscr{R} \gg r_0$, an assumption that we can check.  With these considerations, we find for the total angular momentum and mass

\begin{equation}
\mathscr{L} = \frac{2\pi^{3/2}\rho_0\sqrt{{a}GM}}{r_0^{\,-q}}\frac{\Gamma(\alpha+3/2)}{\Gamma(\alpha+2)}\frac{\mathscr{R}^{\,-q+7/2}}{-q+7/2}, \label{goodL}
\end{equation}
\begin{equation}
\mathscr{M} = \frac{2\pi^{3/2}\rho_0}{r_0^{\,-q}}\frac{\Gamma(\alpha+1)}{\Gamma(\alpha+3/2)}\frac{\mathscr{R}^{\,-q+3}}{-q+3}. \label{goodM}
\end{equation}
Solving for the radius of the disk in terms of the mass of the disk and the mass of the black hole, we find

\begin{equation}
\mathscr{R} = \bigg{(}\frac{y\kappa\beta\sqrt{a}(3-q)}{4\pi{c}}\mathscr{M}\sqrt{GM_h}\bigg{)}^{2/5} \label{mathscrR} \end{equation}
\begin{equation}
\simeq 9\times10^{14}\bigg{(}\frac{\mathscr{M}}{M_{\astrosun}}\bigg{)}^{2/5}\bigg{(}\frac{M_h}{10^6M_{\astrosun}}\bigg{)}^{1/5} \text{ cm}.
\end{equation}
This relation yields $\mathscr{R} \approx 10^3\,r_s$, $r_s$ being the Schwarzschild radius of the black hole, for $M_h = 10^6M_{\astrosun}$ and $\mathscr{M} = 1M_{\astrosun}$. Because we expect that $r_0 \approx few\times r_s$, we see that neglecting the lower bound in the integrations of $\mathscr{M}$ and $\mathscr{L}$ was justified.

Our goal is to use the total mass and angular momentum, calculable from initial conditions, to determine $q$. This value will then inform us of how a larger progenitor star, a larger black hole, or more angular momentum will influence how steeply the density or pressure falls off with distance from the hole.  By performing a bit of algebra, we can rearrange equations \eqref{goodR}, \eqref{goodL}, and \eqref{goodM} to yield

\begin{multline}
f(\mathscr{M}, \mathscr{L}, M_h) \equiv \bigg{(}\frac{y\kappa}{4\pi{c}}\bigg{)}^{1/6}\frac{\mathscr{M}\sqrt{GM_h}}{\mathscr{L}^{5/6}} = \\
\frac{\Gamma(\alpha+1)^{5/6}\Gamma(\alpha+2)^{5/6}}{\beta^{1/6}a^{1/2}\Gamma(\alpha+3/2)^{5/3}}\frac{(7/2-q)^{5/6}}{3-q}. \label{neq}
\end{multline}
The left-hand side of this expression, denoted $f(\mathscr{M},\mathscr{L},M_h)$, depends only on the total mass of the disk, the total angular momentum of the disk, and the black hole mass (in addition to a few physical constants; note that its dependence on $y$, the parameter we introduced to describe the efficiency of convection, is to the 1/6th power, and therefore only affects our answers very weakly). The right-hand side, on the other hand, is only a function of $q$, which we could in principle invert to isolate $q$ itself. The gross properties of the progenitor star and the black hole therefore determine the density, pressure, and angular momentum profiles of the fallback disks associated with super-Eddington TDEs.

In order to calculate $q$ for a given TDE, we need to parametrize the total mass and angular momentum in terms of those values appropriate to a certain event, both of which will depend on the progenitor star. In order to be tidally disrupted, the star must pass within the tidal radius $r_t \approx R_*(M_h/M_*)^{1/3}$ of the black hole, where $R_*$ is the stellar radius and $M_*$ is its mass (the precise point of disruption clearly depends on the details of the stellar composition, rotation, and other complications, but numerical results indicate that the true location does not vary from that given by more than a factor of $ \sim 1.5$ for realistic interiors; \citealt{iva01}). Due to the tidal force on the star and the tidal potential, nearly half of the stellar debris is ejected from the black hole on hyperbolic orbits \citep{lac82}.  The other half remains bound to the SMBH.  The initial mass of the disk should therefore be on the order of $\mathscr{M} \approx M_*/2$, though the actual amount should be slightly less than this when we account for material that has already been accreted and the still-raining-down debris outside $\mathscr{R}$ (see section 4). At the tidal radius, conservation of energy dictates that the star has a velocity of $v_* = \sqrt{2GM_h/r_t}$, and hence the disk material has a total angular momentum of $\mathscr{L} \approx M_*\sqrt{GM_hR_*/2}(M_h/M_*)^{1/6}$ (again, this is a slight overestimate). By parametrizing the mass and angular momentum as such, equation \eqref{neq} becomes

\begin{equation}
5\,y^{\,1/6}\frac{M_{*\astrosun}^{\,11/36}}{M_6^{\,1/18}R_{*\astrosun}^{\,5/12}} = \frac{\Gamma(\alpha+1)^{5/6}\,\Gamma(\alpha+2)^{5/6}}{\beta^{1/6}a^{1/2}\,\Gamma(\alpha+3/2)^{5/3}}\frac{(7/2-q)^{5/6}}{3-q}. \label{neq2}
\end{equation}
Here $M_{*\astrosun}$ is the progenitor's mass in units of solar masses, $R_{*\astrosun}$ is its radius in units of solar radii, and $M_6$ is the black hole mass in units of $10^6M_{\astrosun}$. Interestingly, the left-hand side is virtually independent of the black hole mass, meaning that the density and pressure distributions of TDE fallback disks are almost exclusively determined by the progenitor star.

\begin{figure}[h] 
   \centering
   \includegraphics[width=3.5in]{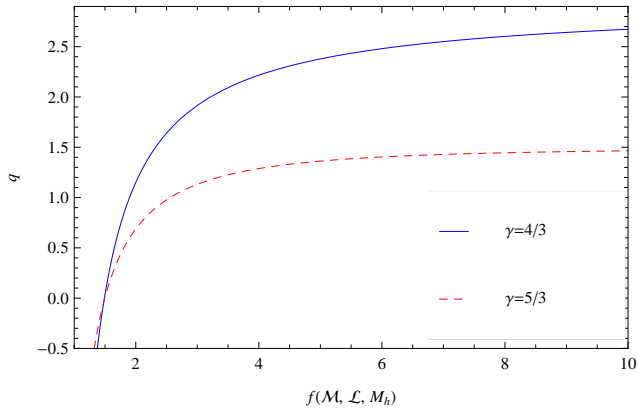} 
   \caption{The value of $q$ obtained for $\gamma = 4/3$ (blue, solid) and $\gamma = 5/3$ (red, dashed) as we vary the left-hand side of equation \eqref{neq}.}
   \label{fig:ngamm}
\end{figure}

Figure \ref{fig:ngamm} illustrates the value of $q$ obtained as we vary the function $f(\mathscr{M},\mathscr{L},M_h)$. As we increase $f(\mathscr{M},\mathscr{L},M_h)$, the value of $q$ approaches $q \rightarrow 1/(\gamma-1)$.  By analyzing equation \eqref{alphaeq}, we can show that $\alpha=0$ for this value of $q$; recalling that $\rho \propto (\sin^2\theta)^{\alpha}$, we see that the flow is spherically symmetric. This result makes sense when we realize that, in order for $f(\mathscr{M},\mathscr{L},M_h)$ to approach large values, the angular momentum must be very small.

The value of $q$ rapidly decreases as we decrease $f$. Recalling our lower limit on $q$, namely that $q > 1/2$, we see that there is a lower limit on the value of the left-hand side of \eqref{neq} which is, after further investigation of Figure \ref{fig:ngamm}, nearly independent of the adiabatic index. Numerically we find

\begin{equation}
y^{\,1/6}\frac{\mathscr{M}\sqrt{GM}}{\mathscr{L}^{5/6}} \gtrsim \begin{cases}
179 \text{ for } \gamma = 5/3 \\
163 \text{ for } \gamma = 4/3
\end{cases},
\end{equation}
where we have used the opacity for Thomson scattering and the units are cgs. If one violates these lower bounds, our model ceases to describe the disk adequately. We can show that, if the inequality is not satisfied, then $\mathscr{R} < R_c$, where $R_c = \mathscr{L}^2/(GM_h\mathscr{M}^2)$ is the circularization radius, which is obtained by balancing gravity and the centrifugal force. When $f(\mathscr{M},\mathscr{L},M_h)$, defined in equation \eqref{neq}, falls below this critical value, pressure forces play a negligible role in the dynamics of the system, and the disk becomes thin.  It is therefore no surprise that our model breaks down in this limit. 

The total mass and angular momentum of the disk thus determine the large scale properties of the envelope, which in turn determine the density and pressure profiles. Extrapolating these profiles to the region of the disk near the black hole, we can estimate conditions in the vicinity of the innermost stable circular orbit, which then quantify the accretion rate and rate of energy generation. Because any further absorption of energy would lead to a positive Bernoulli parameter and unbind the envelope (see Appendix B for notions concerning non-zero Bernoulli parameter), we conjecture that the accretion energy must escape through the funnel in the form of a fast jet. 

We would like to be able to say something about the properties of this jet.  Also, the accretion of the black hole and the continual fallback of material outside the envelope (at a rate {roughly proportional to} $t^{-5/3}$ {for later times}) are changing the mass and angular momentum of the system; the values of $q$ resulting from this section should therefore be interpreted as initial, or bulk, parameters. By modeling the mass and angular momentum of the disk in a time-dependent manner, we will be able to gain some insight into possible observational diagnostics one could use to infer the presence of a ZEBRA flow.

\section{Jet properties and temporal evolution}
\subsection{Inner regions of the accretion disk}
In the above analysis we assumed that ignoring the inner regions of the accretion disk, where general relativistic effects become important, was permissible. In those regions, however, we know that the angular momentum must exceed its Keplerian value, i.e., that with $\phi = -GM_h/r$, to account for the stronger gravitational acceleration. This excess of angular momentum at smaller radii and its interplay with the pressure gradient could, in principle, significantly alter the flow at larger radii and change our results. 

Models that investigated the inner regions of thick disks around black holes, termed ``Polish doughnuts", were developed in the late 1970's and early 80's, and research along these lines continues to the present day \citep{abr78,koz78,pac80,jar80,pac82,kom06,qia09,abr13,pug13}. In these models, authors assume ad hoc forms for both the specific entropy and angular momentum, and from these functional forms one may infer the pressure and density from the relativistic energy and momentum conservation equations. The portion of the ZEBRA envelope closest to the black hole should in many respects resemble a Polish doughnut, with the extra constraint that the flow has zero {Bernoulli function}. Instead of pursuing the lines followed by many authors in examining the consequences of the relativistic conservation equations, we will follow a slightly different route which incorporates our model.

To analyze the specifics of the flow near the black hole and its impact on the outer regions of the envelope, we will restrict our attention to the case where the space-time metric is that of Schwarzschild. With this assumption, we then replace the standard point-mass potential with the ``pseudo-Newtonian" potential of \citet{pac80}, so $\phi \rightarrow -GM_h/(r-r_s)$, where $r_s = 2GM_h/c^2$ is the Schwarzschild radius. While this potential tends to produce inaccurate numbers for some quantities \citep{tej13}, its prediction of the innermost stable circular orbit and the marginally bound orbit suffice for our treatment. 

By manipulating the momentum and Bernoulli equations with this potential, we can show that the most general form of the angular momentum must satisfy $\ell^2(r,\theta) = \ell^2(\phi\,{r^2}\sin^2\theta)$, i.e., the angular momentum is only a function of the combination $\phi\,r^2\sin^2\theta$ (note that this result is consistent with equation \eqref{angmom} in which a specific basis set for the functions is used). In the self-similar limit, we showed that $\ell^2 = a\phi\,r^2\sin^2\theta$. However, in addition to approaching the self-similar value in the $r \rightarrow \infty$ limit, the angular momentum must also match that of the psuedo-Newtonian distribution (that with the Paczy\'nski-Wiita potential but without a pressure gradient) at some inner radius where the pressure gradient goes to zero. Because the self-similar solution will not necessarily satisfy the second condition, we must search for non-self-similar distributions. The specific form we will adopt is $\ell^2 = A + D\,GM_h\,r^2/(r-r_s)$, where $A$ and $D$ are constants and we are restricting our attention to the equatorial plane. In general the angular momentum distribution could be more complicated. However, it must monotonically increase with radius throughout -- a decrease in the specific angular momentum with radius is highly unstable to convection \citep{gol67,seg75}, unless it is accompanied by a strong increase in entropy, which is unlikely. It must also approach the self-similar solution in the asymptotic limit.  The previous form is the simplest that satisfies both of these criteria.

Requiring that the angular momentum approach its self-similar value for large $r$ yields $D = a$, where $a$ is given by equation \eqref{aeq}. With our specific form for the angular momentum, we can manipulate the momentum equations to find exact expressions for both the density and the pressure. Setting the pressure gradient equal to zero at some radius $r_m$ where $\ell^2(r_m) = \ell^{2}_{PN}(r_m)$, where $\ell_{PN}^2 = 2GM_hr(r/(r-r_s))^2$ is the pseudo-Newtonian angular momentum, yields $A = 8GM_hr_s(1-a/2)$ and $r_m = 2\,r_s$, which is the marginally bound orbit. Our solution for the self-consistent angular momentum is thus

\begin{equation}
\ell^2 = GM_h\bigg{(}4r_s(2-a)+\frac{ar^2}{r-r_s}\bigg{)}. \label{ellexact}
\end{equation}

\begin{figure}[htbp] 
   \centering
   \includegraphics[width=3.5in]{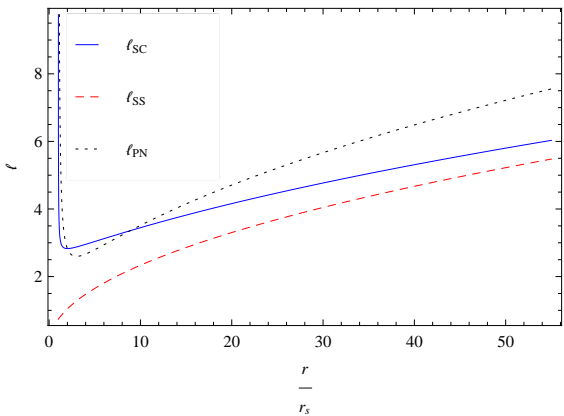} 
   \caption{The self-consistent model (equation \eqref{ellexact}) for the angular momentum (blue, solid),  the self-similar model (red, dashed), and the psuedo-Newtonian distribution (black, dotted). Here we've set $q = 1.5$, $\gamma = 4/3$, the abscissa is in units of Schwarzschild radii, and the angular momentum is normalized by $\sqrt{GM_hr_s}$.}
   \label{fig:ellplots}
\end{figure}
Figure \ref{fig:ellplots} illustrates three different angular momentum distributions for a given set of parameters: that given in equation \eqref{ellexact}, the self-similar solution, and the pseudo-Newtonian distribution. As we can see, the self-consistent model flattens out in the inner region to exceed both the pseudo-Newtonian distribution and the self-similar angular momentum. 

It may seem like this excess of specific angular momentum could alter significantly our estimates of $q$, $\mathscr{R}$, and other properties of the envelope. However, the two conserved quantities in a tidal disruption event are the \emph{total} angular momentum and \emph{total} mass, and it is not clear how much these differ from those in the self-similar limit. Therefore, to answer whether or not this modified potential truly affects our results, we must also determine how the density varies in the self-consistent limit. After evaluating the density, we can form the integrals $\int^{r}4\pi{r^2}\rho\,dr$ and $\int^{r}4\pi{r^2}\ell\,\rho\,dr$ to obtain the enclosed mass and angular momentum as functions of $r$, respectively (the lower bound will not significantly affect the result in either case). By comparing these functions to their analogs in the self-similar limit, we can assess how significantly the pseudo-Newtonian potential affects our conclusions.

\begin{figure}[htbp] 
   \centering
   \includegraphics[width=3.5in]{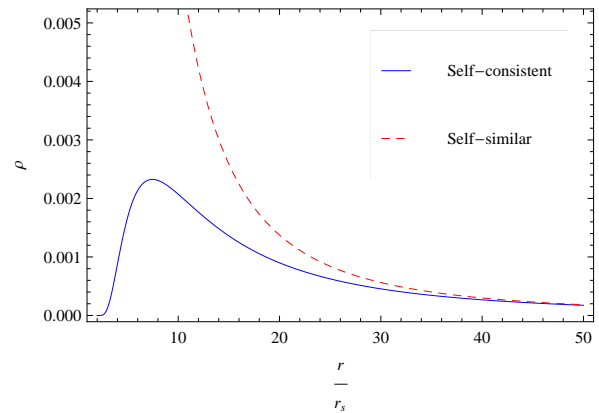} 
   \caption{Solutions for the density in the non-self-similar (blue, solid) and self-similar (red, dashed) limits. Here $q = 1.5$, $\gamma = 4/3$, the x-axis is in units of Schwarzschild radii, and the density is measured in {units such that $\rho_0\,(r_s/r_0)^{-q} = 1$, i.e., the red, dashed curve is simply $(r/r_s)^{-2.2}$}.}
   \label{fig:rhoplots}
\end{figure}

\begin{figure}[htbp] 
   \centering
   \includegraphics[width=3.5in]{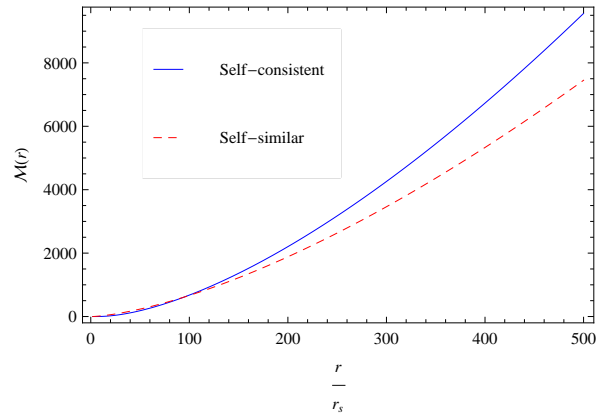} 
   \caption{The mass contained within $r$ for the angular momentum given in equation \eqref{ellexact} (blue, solid) and that for the self-similar model (red, dashed). The parameters are the same as those in Figure \ref{fig:rhoplots}{, with the same normalization for the density}.}
   \label{fig:masses}
\end{figure}

\begin{figure}[h] 
   \centering
   \includegraphics[width=3.5in]{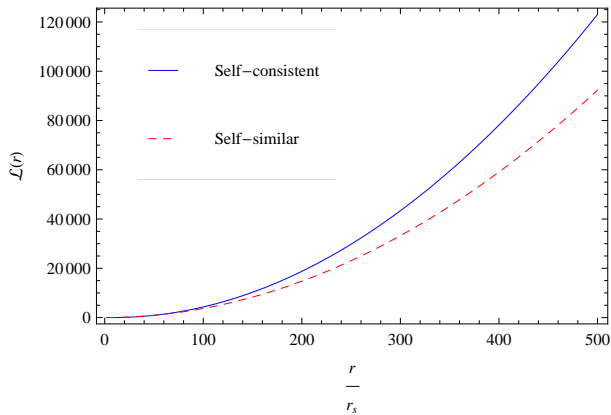} 
   \caption{The angular momentum contained within $r$ for the self-consistent model (blue, solid) and the self-similar solution (red, dashed). The parameters{ and normalization} are the same as those in Figure \ref{fig:rhoplots}.}
   \label{fig:angmoms}
\end{figure}

Figure \ref{fig:rhoplots} displays the density for both the self-consistent (using equation \eqref{ellexact} for the angular momentum) and self-similar solutions. At small radii strong gravity reduces the density from that in the self-similar limit, while at large radii they are indistinguishable. Figure \ref{fig:masses} illustrates the total mass contained within $r$ for the self-similar and non-self-similar models, while Figure \ref{fig:angmoms} shows the total angular momentum contained within $r$ for both models. The self-consistent model predicts an increased amount of mass and angular momentum at larger radii, which is due to a slight increase in its density, relative to the self-similar solution, at intermediate radii. This means that the use of the Pascy\'nski-Wiita potential makes a more compact ZEBRA (in other words, we would enclose the same amount of mass and angular momentum, fixed by the TDE, at a smaller radius). However, we do not believe that this alteration will change our results {much, as the physics is largely dictated by the ratio of the total angular momentum to mass, which is conserved from the TDE}. Thus, while relativity can alter significantly the behavior of the density, pressure, and angular momentum at small radii, its effects on the bulk properties of the ZEBRA are minimal. We therefore expect that its inclusion in our models will not significantly change the results.

If the Bernoulli parameter were very small (compared to $GM_h/r$) all the way to the black hole, the gas would release little energy in the form of a jet. However, we expect that this is not realistic, and that fluctuations in the inner part of the flow will lead to the root-mean-squared value of the binding energy being some significant fraction of that for the innermost stable circular orbit (ISCO) ($=6GM_h/c^2$ for a Schwarzschild black hole). To quantify this statement, recall that, for a non-steady flow, an additional term must be added to the Bernoulli function:

\begin{equation}
B \rightarrow B + \int\frac{\partial\bold{v}}{\partial{t}}\,\cdot{d}\bold{r},
\end{equation}
where the integral is taken over the flow line of the fluid element. In terms of scalings, this term is on the order

\begin{equation}
\int\frac{\partial\bold{v}}{\partial{t}}\,\cdot{d}\bold{r} \sim \frac{\Delta{v}\,r}{\tau_d}.
\end{equation}
Here $\Delta{v}$ is the change in the velocity over the dynamical timescale, $\tau_d$, and $r$ is the radius at which we are considering the (predominantly circular) flow line. Using $\tau_d \sim r/v$ and $\Delta{v} \sim v$, we find $\Delta{v}\,r/\tau_d \sim v^2$ (there is also a term, of order $\sim v^2$, that arises from the inclusion of the $\bold{v}\cdot\nabla\bold{v}$ term in the momentum equation; however, because it is of the same order as the time derivative, it suffices to consider only this term). At the ISCO and regions interior to that radius, this fluctuation term can reach substantial fractions of $c^2$. Because the change in velocity can be either positive or negative, the average will cancel out over the whole flow. The root-mean-squared of the additional term will not cancel, however, and will lead to fluctuations that lead to both positive and negative Bernoulli parameter. The fluctuations to the negative side give rise to bound flows, which then release that energy in the form of a very energetic jet.

As in the case of ADIOS models \citep{bla04}, the mechanism responsible for this energy dissipation is not specified --- its existence and nature will have to be verified later.  Once this energy is injected, however, its escape will most likely be in the form of a jet due to the rotational funnel.  Inside the radius at which the material becomes bound, our previous model breaks down. Because this region constitutes a minute fraction of the disk, we do not expect that its presence will have much of an impact on our results.

\subsection{Accretion rate and jet power}
We derived the density, pressure, and angular momentum distributions for ZEBRA envelopes under the assumption that the poloidal and radial velocities were significantly less than that in the azimuthal direction. While this proposition is upheld in the bulk of the flow, it must break down in regions near the SMBH, specifically in those regions where relativity prevents the existence of stable circular orbits. It follows that the radial velocity in this region should be appropriate to that of gravitational free-fall, or $v_r \sim \sqrt{GM_h/r}$.

To make contact with the self-similar region of the flow, we will take $r_0$ to be the radius at which gravitational infall becomes substantial, and we have $v_{r_0} = \delta\sqrt{GM_h/r_0}$, where $\delta$ is a number less than one. With this assumption, the mass accretion rate onto the black hole should be on the order of that in the spherically symmetric regime, or $\dot{M}_{acc} = 4\pi{r_0^2}\rho{v_{r_0}}$. Plugging in our expressions for relevant quantities, we find

\begin{equation}
\dot{M}_{acc} = 4\pi{\delta}\rho_0\sqrt{GM_h}r_0^{3/2}.
\end{equation}
We can solve for $\rho_0$ in terms of the inner radius, the total mass of the disk and the mass of the black hole. Doing so and putting the expression into the equation for $\dot{M}_{acc}$ yields

\begin{multline}
\dot{M}_{acc} = \delta\,\chi^{3/2-q}\mathscr{M}\sqrt{GM_h}\bigg{(}\frac{2GM_h}{c^2}\bigg{)}^{3/2-q}\\ \times\bigg{(}\frac{y\sigma_T}{4\pi{c}m_p}\mathscr{M}\sqrt{GM_h}\bigg{)}^{-\frac{2}{5}(3-q)}h(q), \label{mdot}
\end{multline}
where

\begin{equation}
h(q) \equiv \frac{2}{\sqrt{\pi}}\frac{\Gamma(\alpha+3/2)(3-q)}{\Gamma(\alpha+1)}\bigg{(}(3-q)\beta\sqrt{a}\bigg{)}^{-\frac{2}{5}(3-q)},
\end{equation}
which is a function only of $q${ and $\gamma$}. We have also parametrized the inner radius in terms of the Schwarzschild radius, viz. $r_0 = \chi\,2GM_h/c^2$.

The jet luminosity is given by $L_j = \epsilon\dot{M}_{acc}c^2$, where $\epsilon$ is the accretion efficiency of the black hole. To arrive at this result it was necessary to introduce a number of factors that relate to our uncertainty of the details of the flow, namely $\epsilon$, the radiative efficiency, $\chi$, the inner edge of the disk, $\delta$, the fraction of free-fall of the velocity, and $y$, the advective efficiency. However, we expect that $\epsilon$, $\delta$, and $y$ are somewhere in the range of $0.01-1.0$, and we know that $\chi$ should be on the order of a few (strictly, this value and the radiative efficiency depend on the spin of the hole and its orientation relative to the disk). Therefore, although we have a number of unknowns, their range in parameter space is rather small. Explicitly we find for the jet power

\begin{multline}
L_j = \mu\,\mathscr{M}c^2\sqrt{GM_h}\bigg{(}\frac{2GM_h}{c^2}\bigg{)}^{3/2-q}\\ \times\bigg{(}\frac{\sigma_T}{4\pi{c}m_p}\mathscr{M}\sqrt{GM_h}\bigg{)}^{-\frac{2}{5}(3-q)}h(q),
\end{multline}
where we set $\mu \equiv \epsilon\,\delta\,\chi^{3/2-q}\,y^{-2(3-q)/5}$ for compactness.  For a solar progenitor, a million-solar-mass black hole, $y = 0.5$, $\delta = 0.05$, and $\epsilon = 0.1$, we find  $L_j \approx 5\times10^{47}\text{ erg s}^{-1} \approx 4\times10^3L_{Edd}$ for the jet luminosity, where $L_{Edd} = 4\pi{G}cM_hm_p/\sigma_T$ is the Eddington luminosity of the black hole assuming ionized hydrogen (here we have solved equation \eqref{neq2} to determine the value of $q$, which, for these numbers, is $q \approx 2.4$). 

\subsection{Time-dependent analysis}
{In order to be tidally disrupted, the stellar progenitor must pass within a pericenter distance of $r_p = x\,r_t$, where $r_t = R_*(M_h/M_*)^{1/3}$ is the tidal radius and $x$ is a number that is less than or about one. {Here we will restrict our attention to the case where $x=1$. Our motivation for doing so is that other authors have shown, using hydrodynamical simulations, that the complexities of the encounter for smaller and larger $x$ render an analytical treatment insufficient for describing the physics of the TDE \citep{gui13}. Because running a numerical simulation to determine the exact feeding rate to the ZEBRA is outside the scope of this paper, we will only consider those disruptions which occur exactly at the tidal disruption radius and maintain that the analytical approach is accurate enough for our purposes.} 

After} the star is disrupted, the most tightly bound material is placed on an orbit with semi-major axis $R_i = (R_*/2)(M_h/M_*)^{2/3}$ (see section 1 for a derivation). Because the point of disruption occurs {at} the tidal {disruption} radius, the eccentricity of this orbit is very large. Other less-bound gas parcels (those with larger semi-major axes) are thus on nearly-parabolic orbits as they recede from the hole.  The initial configuration of the tidally-stripped material {that is going to fall back} is therefore a thin, highly-elliptical disk, confined roughly to the plane occupied by the disrupted star.

When the innermost gas undergoes one complete orbit, shock heating and other effects (see Introduction) begin to circularize the orbits and alter the structure of the debris disk. After a certain amount of time has passed, on the order of a few orbits of the innermost material, the heating causes the disk to puff up into a spheroid of radius $\mathscr{R}_0$ -- this is the ZEBRA. In the previous sections we assumed, for simplicity, that the entire mass that is bound to the black hole (nearly half the stellar progenitor) comprised the ZEBRA. However, because tidally-stripped material is continually falling back onto the accretion region, it is not clear that this assumption is valid.

To determine how much mass is contained in the initial ZEBRA {and the rate at which material is falling back onto the accretion region, we will pursue a line of analysis similar to that in \citet{lod09}, and} consider the star at the time of disruption. At this point in time, the center of mass is at the tidal radius, and, assuming the star is on a parabolic orbit, the binding energy of the center of mass is zero. Denote the position of a gas parcel contained in the star by $r_g = r_t-\eta{R_*}$; $\eta = 1$ is the edge of the star closest to the hole, $\eta = -1$ is that farthest from the hole, and we are restricting our attention to the plane of the orbit. The specific gravitational energy of a gas parcel is then given by

\begin{equation}
\epsilon_p = \frac{GM_h}{r_t}-\frac{GM_h}{r_t-\eta{R_*}}
\end{equation}
\begin{equation}
\simeq -\frac{GM_h{R_*}}{r_t^2}\eta,
\end{equation}
where in the final line we approximated the tidal radius as being much greater than the stellar radius, valid for the supermassive black holes we are considering. After disruption, the gas parcels fly apart, cooling adiabatically and occupying roughly Keplerian orbits. The energy-period relation for Keplerian orbits yields the semi-major axes of these orbits:

\begin{equation}
R_p = \frac{R_*}{2\eta}\bigg{(}\frac{M_h}{M_*}\bigg{)}^{2/3} = \bigg{(}\frac{\sqrt{GM_h}}{2\pi}\bigg{)}^{2/3}t^{2/3} \label{semieq},
\end{equation}
{where $t$ is the fallback time.} Note that, by inserting $\eta = 1$, this expression reproduces the correct orbit for the most tightly bound debris.

\begin{figure}[htbp] 
   \centering
   \includegraphics[width=3.5in]{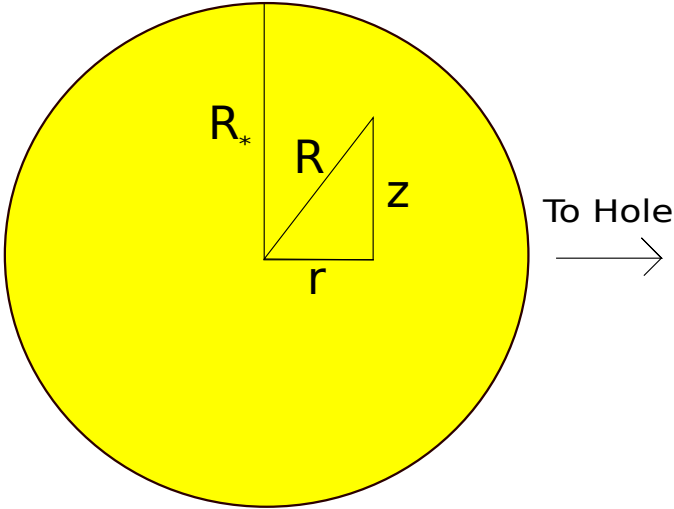} 
   \caption{A schematic of the star at the time of disruption to illustrate the geometry. Note that $r = \eta{R_*}$.}
   \label{fig:disrupted_star}
\end{figure}

{The rate at which material returns to pericenter is found by using the chain rule, specifically $dM/dt = (dM/d\eta)(d\eta/dt)$. From equation \eqref{semieq}, we can readily determine $d\eta/dt$. To calculate $dM/d\eta$, we will assume that the stellar progenitor is well-approximated by a polytropic equation of state; a number of authors have shown that the equation of state of the star has important consequences for the rate of return of material, and so it is not adequate simply to consider a constant-density profile \citep{lod09,mac12,bog13, gui13}. In this case, $\rho(R) = \lambda\,\theta^{\,1/(\gamma_*-1)}$, where $\lambda$ is the density at the center of the star, $\gamma_*$ is the polytropic index of the gas that comprises the star, and $\theta(R)$ is the solution to the Lane-Emden equation. $R$ is spherical distance measured from the center of the stellar object. We will parametrize the location of a gas parcel within the star in terms of the variables $R$, $r$ and $z$, where $r$ is the distance from the center of the star in the plane of the orbit and $z$ is the distance perpendicular from the plane of the orbit (see Figure \ref{fig:disrupted_star} for clarification). Using the fact that $dM = \rho\,{dV}$, where $dV = 2\pi{z\,dz\,dr}$ is the volume element, and making simple geometric substitutions, we can show that}

\begin{equation}
\frac{dM}{d\eta} = 2\pi\int_{\eta{R_*}}^{R_*}\rho\,R\,dR
\end{equation}
\begin{equation}
=\frac{M_*\xi_1}{2}\frac{\int_{\eta\xi_1}^{\xi_1}\theta^{\frac{1}{\gamma_*-1}}\xi\,d\xi}{\int_0^{\xi_1}\theta^{\frac{1}{\gamma_*-1}}\xi^2\,d\xi},
\end{equation}
{where $\xi$ is the dimensionless radius defined through the Lane-Emden equation and $\xi_1$ is the first root of $\theta(\xi)$ (see, e.g., \citealt{han04}). Putting everything together, we obtain for the mass rate of return}

\begin{multline}
\frac{dM}{dt} = \frac{M_*R_*\xi_1}{6}\bigg{(}\frac{2\pi{M_h}}{M_*\sqrt{GM_h}}\bigg{)}^{2/3}t^{-5/3} \\ \times\frac{\int_{\eta\xi_1}^{\xi_1}\theta^{\frac{1}{\gamma_*-1}}\xi\,d\xi}{\int_0^{\xi_1}\theta^{\frac{1}{\gamma_*-1}}\xi^2\,d\xi} = \dot{M}_{fb} \label{maccrate}.
\end{multline}
{This expression gives the rate at which material is drained from the tidally-disrupted debris cloud and added to the ZEBRA. Note that equation \eqref{maccrate} only holds for $\eta < 1$, or for times $t > t_r$, where}

\begin{equation}
t_r = \bigg{(}\frac{R_*}{2}\bigg{)}^{3/2}\frac{2\pi{M_h}}{M_*\sqrt{GM_h}} \label{tr}
\end{equation}
{is the time taken for the innermost material to undergo one complete orbit. Noting that the original mass contained in the bound tidally-disrupted material is roughly $M_*/2$, we can write an expression for the remaining mass that is still raining down onto the ZEBRA after a time $t$:}

\begin{equation}
M_{fb}(t) = \frac{M_*}{2} - \int_{t_r}^t\frac{dM}{dt'}dt' \label{fallbackmass},
\end{equation}
{where $dM/dt'$ is given by expression \eqref{maccrate} with $t \rightarrow t'$.}

The jet and black hole are also extracting angular momentum from the disk. However, it is necessary for the disk material to transport a large amount of its angular momentum outwards, via viscous or magnetic processes (neither of which we have attempted to include in this model) in order to be accreted by the black hole. The angular momentum of the disk material is thus nearly unaffected by the presence of the hole.  However, as we noted previously, natal stellar material is still falling back onto the envelope. As we have assumed that hydrodynamic effects have only influenced the particles in the region of the ZEBRA envelope, this material still approximately retains its specific angular momentum from the time of disruption, adding this angular momentum to the disk as it falls back. {Using equation \eqref{fallbackmass}} , we find for the total angular momentum as a function of time

\begin{equation}
\mathscr{L} = \sqrt{2GM_hR_*}\bigg{(}\frac{M_h}{M_*}\bigg{)}^{1/6}\bigg{(}\frac{M_*}{2}-M_{fb}(t)\bigg{)}
\end{equation}

After the ZEBRA has inflated to a radius $\mathscr{R}_0$, the envelope will not only lose mass to the black hole at the rate described by equation \eqref{mdot}, but it will also gain mass at the expense of the tidally-stripped debris that is still {falling} back. We can solve for $\mathscr{M}$ in terms of other quantities by rearranging equation \eqref{neq}, {and our differential equation for $q$ is then $\dot{\mathscr{M}} = dM/dt-\dot{M}_{acc}$, where $dM/dt$ is given by equation \eqref{maccrate} and $\dot{M}_{acc}$ by equation \eqref{mdot}. To solve this differential equation, we need an initial value for $q$. This initial condition can be determined by assuming that the ZEBRA takes some time to inflate, at which point it has some mass and angular momentum, which in turn yield an initial $q$. However, the time to inflate depends sensitively on the heating rates and other physical processes, for which we do not have a reliable model.  We also expect that any knowledge of the initial conditions should be lost after a certain amount of time, and that they should only reflect a transient initial behavior. We will therefore leave the initial value of $q$, which we will denote $q_0$, as an unspecified parameter, and only when systems with different $q_0$ converge on a single solution will we consider our models accurate. After numerically integrating the differential equation for $q(t)$, }we can go on to compute $\mathscr{M}(t)$, $\mathscr{L}(t)$ and $L_j(t)$.

\begin{figure}[htbp] 
   \centering
   \includegraphics[width=3.5in]{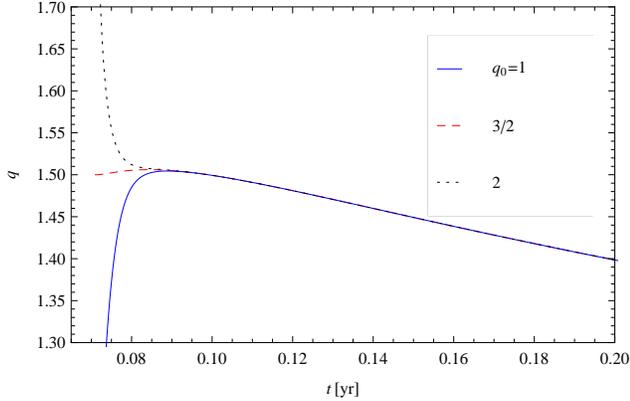} 
   \caption{The solution for $q(t)$ with a solar progenitor ($M_* = 1\,M_{\astrosun}$, $R_* = 1\,R_{\astrosun}$, $\gamma_* = 5/3$), a radiation pressure-dominated gas ($\gamma = 4/3$), $M_h = 10^5M_{\astrosun}$, $y = 1$, $\delta = 0.05$, $\chi = 5$, and three different $q_0$, indicated by the legend. As one can see, the initial conditions quickly become irrelevant to the long-term behavior of the solutions.}
   \label{fig:qplotsq0}
\end{figure}

\begin{figure}[htbp] 
   \centering
   \includegraphics[width=3.5in]{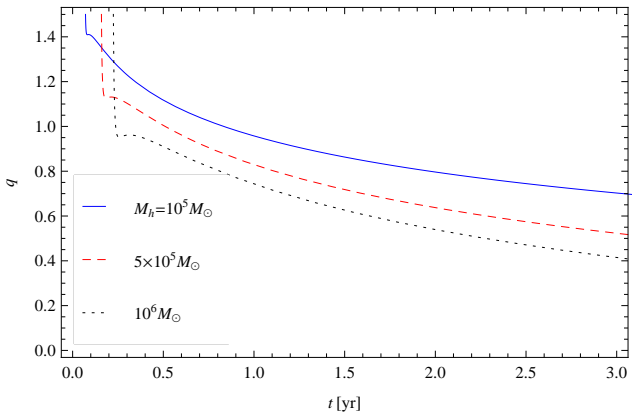} 
   \caption{$q(t)$ for a radiation-dominated gas ($\gamma = 4/3$) and a variety of black hole masses.  Here $y=0.5$, $\delta = 0.05$, $\chi = 5$, $q_0 = 3/2$, $\gamma_* = 5/3$, $M_* = 1M_{\astrosun}$, and $R_* = 1R_{\astrosun}$. The legend displays the black hole mass in units of solar masses. We see that initially $q$ {falls off very rapidly, which is a consequence of initial conditions. However, the initial conditions stop having a major effect early in the evolution of the system, and $q$ then decreases less rapidly.}}
   \label{fig:nplotsmbh}
\end{figure}

\begin{figure}[htbp] 
   \centering
   \includegraphics[width=3.5in]{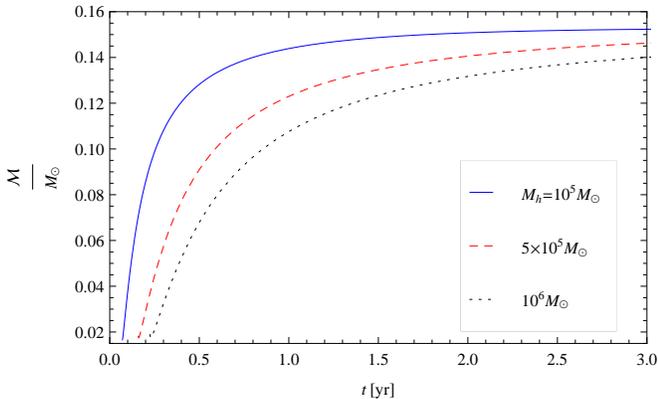} 
   \caption{The mass contained in the ZEBRA as a function of time for the same parameters as those in Figure \ref{fig:nplotsmbh}. The mass quickly {increases initially, owing to the fact that the fallback rate exceeds the accretion rate.} However, as {both rates decrease for later times}, the mass levels off to a nearly constant value.}
   \label{fig:massplotsmbh}
\end{figure}

\begin{figure}[htbp] 
   \centering
   \includegraphics[width=3.5in]{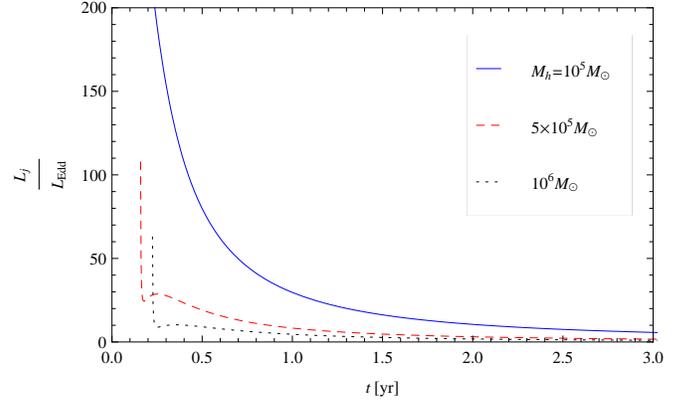} 
   \caption{The jet luminosity, normalized to the Eddington luminosity of the black hole, for the same parameters as those in Figure \ref{fig:nplotsmbh} with an efficiency $\epsilon = 0.1$. Initially the luminosity is very super-Eddington, decaying to only mildly super-Eddington at later times.}
   \label{fig:ljplotsmbh}
\end{figure}

{Figure \ref{fig:qplotsq0} demonstrates how $q(t)$ changes as we alter $q_0$ for a solar progenitor, a $10^5M_{\astrosun}$ black hole, and a number of other input values.  As expected, the initial conditions strongly influence the behavior of $q(t)$ for early times, but, after about $0.1$ years, which is about $2.5\,t_r$ for this configuration, the different values become indistinguishable. We can thus say with confidence that after this time our models represent the fully-inflated ZEBRA. This timescale, namely a few revolutions of the innermost material, is also consistent with our expectations concerning the amount of time needed for the shock heating to add enough energy to the system.} Figure \ref{fig:nplotsmbh} shows $q(t)$ for various black hole masses and a set of fiducial parameters. {The initial conditions cause $q(t)$ to decrease rapidly. However, the knowledge of such initial conditions is quickly lost, and the system settles into a state in which $q(t)$ decreases less rapidly.}  Figure \ref{fig:massplotsmbh} shows the mass contained in the envelope as a function of time. Because {the fallback rate exceeds the accretion rate, the mass initially increases rapidly. For later times}, the accretion rate and the fallback rate both drop significantly enough to leave a roughly constant mass. Figure \ref{fig:ljplotsmbh} plots the jet luminosity; as we can see, the luminosity predicted by our model is super-Eddington for a significant amount of time, though the amount of time for which that statement is true decreases as the black hole mass increases.  The fact that the jet power is supercritical is a good consistency check on our model. The time at which the accretion rate becomes sub-Eddington is roughly the same time at which $q = 0.5$, where our model begins to break down.

One of the {popularly-cited} hallmarks of a tidal disruption event is that the accretion luminosity is proportional to $t^{-5/3}$. {However, this result only holds for a constant-density star. Our models account for the mass distribution in the original progenitor, placing more mass on orbits with larger semi-major axes, and they are also consistent with the initial rise in the fallback rate. Our accretion rate also} takes into account fluid interactions{, meaning that our black hole accretion rate, and consequently the jet luminosity, does not necessarily mimic exactly the mass fallback rate.} 

\begin{figure}[htbp] 
   \centering
   \includegraphics[width=3.5in]{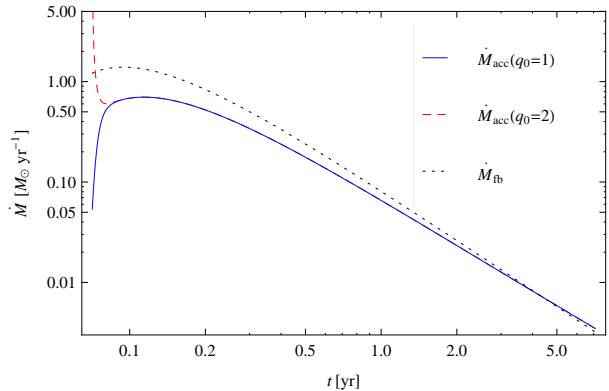} 
   \caption{{The black hole accretion rate, given by equation \eqref{mdot}, in solar masses per year for $q_0 = 1$ (blue, solid curve) and for $q_0 = 2$ (red, dashed curve) to illustrate where the solutions have converged; the parameters are the same as those in Figure \ref{fig:nplotsmbh} and the black hole has a mass $M_h = 10^5M_{\astrosun}$. We have also plotted the fallback rate, shown by the black, dotted curve, to illustrate how the two accretion processes compare. It is apparent from this figure that they match each other closely{, and that after about 4 years, the black hole accretion rate exceeds the fallback rate.}}}
   \label{fig:accvsfb}
\end{figure}

{In Figure \ref{fig:accvsfb} we plot the accretion rate onto the black hole for two different $q_0$ (we chose two different $q_0$ to demonstrate when the solutions converge), given by equation \eqref{mdot}, and the fallback rate onto the accretion region, which is the solution to equation \eqref{maccrate}, for the same set of fiducial parameters as those in Figure \ref{fig:nplotsmbh} and the black hole has a mass $M_h = 10^5M_{\astrosun}$. It is apparent from the figure that the accretion rate onto the black hole follows the fallback rate rather tightly, but there exist notable differences. The first difference is that the accretion rate is less than the fallback rate for the times shown; this finding is consistent with Figure \ref{fig:massplotsmbh}, as the mass is increasing for all times shown. The second is that there exists a temporal lag between the qualitative features shared by the two rates; the most salient example of this characteristic is the difference in the time taken to reach the maximum, which is evident in the figure. Specifically, the fallback rate reaches its maximum at $t \approx 0.094$ years, while the black hole accretion rate peaks at $t \approx 0.11$ years. The third difference is that the accretion rate of the black hole follows a less-steep power law than the fallback rate for later times. By fitting the fallback rate as $\dot{M}_{fb} \propto t^{-m_{fb}}$ between 1 and 2 years, we find that the power law is $m_{fb} \approx 1.63$; by inspecting equation \eqref{maccrate}, we expect that the fallback power law should asymptotically approach $m_{fb} = 5/3$. By fitting the accretion rate for the same amount of time and by the power-law form $\dot{M}_{acc} \propto t^{-m_{acc}}$, we find that $m_{acc} \approx 1.49$. {As one can see in the figure, at a time of about 4 years} the black hole accretion rate exceeds the fallback rate, corresponding to a decrease in the mass contained in the ZEBRA. This result makes sense, as we expect accretion to occur even if there is no fallback of material onto the envelope.}

\begin{figure}[htbp] 
   \centering
   \includegraphics[width=3.5in]{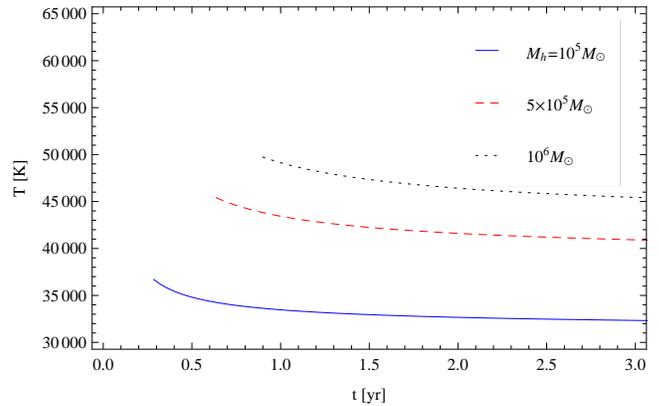} 
   \caption{The effective temperature as a function of time for different black hole masses. The fiducial parameters are the same as those in Figure \ref{fig:nplotsmbh}.}
   \label{fig:teffplotsmbh}
\end{figure}

Another property of the envelope that we can calculate is its effective temperature. As we have argued in section 3, the surface should occur roughly at the trapping radius, and so the temperature is given by

\begin{equation}
T_{} = \bigg{(}\frac{GcM_hm_p}{\sigma_T\sigma_{SB}\mathscr{R}^2}\bigg{)}^{1/4}
\end{equation}
\begin{equation}
\simeq 6.5\times10^4\bigg{(}\frac{M_h}{10^6M_{\astrosun}}\bigg{)}^{1/4}\bigg{(}\frac{\mathscr{R}}{10^{14}\text{cm}}\bigg{)}^{-1/2}\text{ K},
\end{equation}
where $\sigma_{SB} = 5.67\times10^{-5}\text{ cgs}$ is the Stefan-Boltzmann constant. ZEBRA envelopes produced by TDEs thus tend to peak in the far-UV or soft X-ray band. We can also solve for the effective temperature as a function of time, as shown in Figure \ref{fig:teffplotsmbh}.

Because $\mathscr{R}$ is proportional to $\mathscr{M}^{2/5}$ (see equation \eqref{mathscrR}), the temperature {decreases} initially as mass is {gained from the fallback of tidally-stripped material. However, for later times when the black hole accretion rate and the fallback rate both decrease substantially, the temperature remains nearly constant.}

{Owing to the fact that the photons are trapped interior to $\mathscr{R}$, we expect there to be a high degree of coupling between the particles comprising the ZEBRA envelope and the photons produced at the photosphere. The spectrum should therefore be very well-matched by a blackbody distribution. Depending on the temperature of the envelope and the composition of the disrupted star, however, there may also be present a number of absorption and emission features. With these temperatures, electron scattering may also produce a color-corrected spectrum. } 

\subsubsection{Power-law fallback rate}
{As one can see in Figure \ref{fig:accvsfb}, the black hole accretion rate closely matches the fallback rate of the tidally-disrupted material. An interesting question is whether this close equality is always true, or if it just happens to be the case for the specific analytic model that we chose. To analyze the effects of altering the fallback rate, we will let $\dot{M}_{fb}$ scale as a power-law, specifically}

\begin{equation}
\dot{M}_{fb} = M_0(m-1)\,t^{-m} \label{powerlaw},
\end{equation}
{where the proportionality constant has been chosen to be consistent with the fact that at $t_r$, the orbital period of the innermost material, $M_{fb} = M_0$, where $M_0$ is the mass of the material that has yet to be accreted (note, however, that we are not considering the equation accurate until much later than $t_r$, as the fallback rate must first peak and then decline to the power-law decay). From the initial work of \citet{phi89}, it was thought that $m$ should always be on the order of $5/3$. Our Figure \eqref{fig:accvsfb} also indicates that this scaling holds for later times. More recently, however, it has been shown that this power-law decay may not be followed, even for times much later than that at which the peak fallback occurs. In particular, \citet{gui13} demonstrated that variations in the impact parameter, which we defined as $x$, can lead to partial disruptions that, owing to the continued gravitational influence of the surviving stellar core, cause $m$ to deviate significantly from $5/3$. We will therefore leave this quantity as a variable and inquire as to the effects of its variation on the black hole accretion rate. }

{One might expect that $M_0 = M_*/2$, as the TDE leaves roughly half of the progenitor bound to the black hole. However, partial disruptions, which result from grazing encounters with the black hole, leave an intact stellar remnant. In these instances, the total mass bound to the hole is always less than $M_*/2$. We will therefore leave $M_0$ as a free variable, typically on the order of a fraction of $M_*/2$.}

{As we argued previously, the material that comprises the ZEBRA must lose its angular momentum before being accreted by the black hole. By following the same line of reasoning, we can show that the angular momentum contained in the envelope is given by}

\begin{equation}
\mathscr{L} = M_0\sqrt{2GM_hR_*}\bigg{(}\frac{M_h}{M_*}\bigg{)}^{1/6}\bigg{(}1-\bigg{(}\frac{t}{t_r}\bigg{)}^{1-m}\bigg{)}.
\end{equation}
{With this expression, we can go through the same analysis as in the previous subsection and numerically solve for $q(t)$ and all other time-dependent quantities. However, instead of reproducing all of the plots in the previous subsection for different values of $m$, we will concentrate on how the accretion rate onto the black hole compares to the fallback rate, which will inform us of the way in which the jet luminosity relates to the fallback rate.}

{\citet{gui13} demonstrated that $m$ varies between roughly $1.5$ and $2.2$, depending on the value of the impact parameter (see their Figure 7). They also showed that steeper fallback rates follow from shallower impact parameters, as a result of the gravitational influence of the surviving stellar remnant. Consequently, the values of $m$ and $M_0$ are not completely independent. We can show, however, that there exists a nearly-linear scaling between $M_0$ and the magnitude of the black hole accretion rate. Since the fallback rate is also linear in $M_0$, we will simply consider $M_0$ a constant and note that the true value of the accretion rate for a given $m$ may be higher or lower.}

\begin{figure}[htbp] 
   \centering
   \includegraphics[width=3.5in]{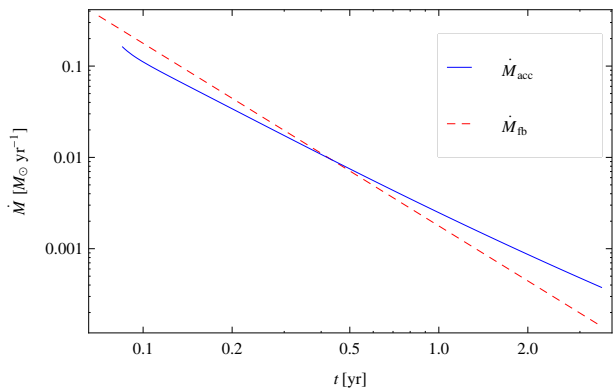} 
   \caption{{The black hole accretion rate (blue, solid curve) and the fallback rate (red, dashed curve) for $M_0 = 0.1\times{M_{\astrosun}}/2$, $M_h = 10^5M_{\astrosun}$, $m = 2$, and otherwise the same parameters as in Figure \ref{fig:nplotsmbh}, plotted on a log-log scale. The accretion rate follows a shallower power-law than the fallback rate, causing the former to exceed the latter for times greater than about 0.5 years.}}
   \label{fig:accvsfb_1}
\end{figure}

{Figure \ref{fig:accvsfb_1} illustrates, on a log-log scale, the accretion rate and the fallback rate for $M_0 = 0.1\times{M_{\astrosun}}/2$, $M_h = 10^5M_{\astrosun}$, $m = 2$, and the parameters adopted in Figure \ref{fig:nplotsmbh}. We see that the accretion rate also follows a power-law decline, but one that is shallower than that for the fallback. Defining $\dot{M}_{acc} \propto t^{-m_{acc}}$, we find, in this case, that $m_{acc} \approx 1.70$.}

\begin{figure}[htbp] 
   \centering
   \includegraphics[width=3.5in]{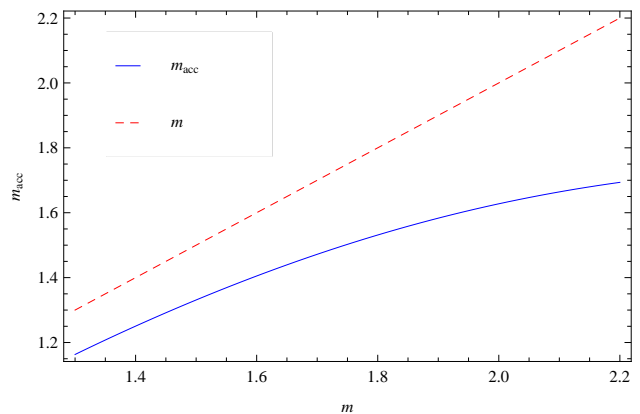} 
   \caption{{The power-law of the accretion rate, which we defined as $m_{acc}$, as a function of the power-law index of the fallback rate is shown by the blue, solid curve. The value of $m_{acc}$ is determined by performing a best-fit to the black hole accretion rate between $t = 0.2$ and $1.5$ years, during which time the accretion rate, for all values of $m$, is well-described by a power-law. We have also plotted $m$ for comparative purposes (red, dashed curve). We see that $m_{acc}$ is always less than $m$. For small values of $m$, the relationship between $m_{acc}$ and $m$ is roughly an offset, linear one. As $m$ becomes larger, however, $m_{acc}$ displays a more nonlinear behavior, and the difference between the two power-law indices becomes larger.} }
   \label{fig:maccvsm}
\end{figure}

{For other power-law fallback rates, a qualitatively similar behavior is exhibited by the accretion rate. In particular, the rate at which mass is accreted by the hole falls off as a power-law, but one that is less steep than the rate at which material impacts the ZEBRA. To illustrate how the value of $m$ affects $m_{acc}$, Figure \ref{fig:maccvsm} shows the value of $m_{acc}$ given the value of $m$. To determine $m_{acc}$, we have performed a best-fit over the timescale of $t = 0.2$ -- $1.5$ years, during which time all of the accretion rates follow power-law decays. As one can see, the relationship between the two power-laws is approximately linear for low values of $m$, becoming increasingly nonlinear as $m$ increases. Thus, while the difference between the two is approximately $\Delta{m} \equiv m-m_{acc} \approx 0.15$ for $m = 1.4$, the disparity becomes $\Delta{m} \approx 0.50$ for $m = 2.2$.}

\subsection{\emph{Swift} J1644+57}
The object \emph{Swift} J1644+57 was found as both a source of X-rays and $\gamma$-rays by the \emph{Swift} satellite, and thought initially to be a gamma-ray burst (GRB) \citep{mar11}. However, the variability and longevity of the source soon proved that such an association was unlikely, and the proximity of the event to the nucleus of a host galaxy at redshift $z = 0.354$ led to the belief that the event was triggered by a TDE \citep{lev11}. By modeling the spatially and temporally coincident radio emission by the interaction of fast-moving ejecta with the circumnuclear environment, it was demonstrated that a mildly relativistic jet was likely generated during the TDE \citep{zau11, met12}. Here we investigate the consistency of our ZEBRA models with the observations of \emph{Swift} J1644+57.

The peak, isotropic X-ray luminosity of \emph{Swift} J1644+57 reached $4\times10^{48}\text{ erg s}^{-1}$, with an average value appropriate to $\sim \text{few}\times10^{47}\text{ erg s}^{-1}$\citep{bur11}. When corrected for beaming effects, we recover a true luminosity of $L_{X} \sim  10^{45}-10^{46}\text{ erg s}^{-1}$ for a jet opening angle of $5^{\circ}$ \citep{blo11}. The mechanism responsible for the generation of the X-rays is still unclear, though its origin is consistent with inverse Compton scattering of photons near the launching point of the jet \citep{mar05,blo11}. It is difficult to constrain directly the mass of the black hole that resides in the host galaxy, but empirical galaxy luminosity relations imply $10^{5}M_{\astrosun} \lesssim M_h \lesssim 10^{6}M_{\astrosun}$ (\citet{sax12} and references therein). The energy generation rate is therefore highly super-Eddington (even if we increase the upper limit of the black hole mass to $10^{7}M_{\astrosun}$).

The prompt evolution of the X-ray emission was highly chaotic. However, after about $ 10 \text{ days}$ from the initial trigger, the flux followed a decline {that was} well-approximated by {a power-law (see {Figure 1 in \citealt{tch13}}).} {Therefore,} if our model is to adequately describe the evolution of \emph{Swift} J1644+57, {the first constraint it must satisfy is that}{ the X-ray flux follow a power-law decline for later times. The exact value of the power-law index is uncertain owing to the large degree of intrinsic scatter in the X-ray data, but it is consistent with the range predicted in \citet{gui13} (also see \citealt{tch13}).}

{The second constraint on our model comes from the fact that the X-ray flux of \emph{Swift} J1644+57 dropped precipitously after about 500 days, }most likely indicating the shut-off of the jet \citep{zau12}.  Because our model requires the accretion rate to be super-Eddington during the jetted phase, our jet luminosity should be roughly the Eddington limit {of the hole} at $t \approx 500\text{ days}$.{ However, because the initial behavior of the X-ray flux was highly chaotic, it is unclear at what point from the time of disruption the \emph{Swift} satellite began observing the source, and the timeline of 500 days is therefore ambiguous.}

{{For the analytical models described in the previous subsection,} the luminosity of the jet depends not only on the black hole mass, but also on a number of other parameters which describe the details of the flow, e.g., $y$, $\delta$, etc. However, the results are largely insensitive to those parameters, and hence we will adopt the values that produced Figures \ref{fig:nplotsmbh} -- \ref{fig:ljplotsmbh}. {Moreover, if the impact parameter differs significantly from $x = 1$, making the power-law model the most valid description of the fallback process, the values of $m$ and $M_0$ must also be incorporated into the model.}

{The flux we observe is altered by the Lorentz factor and opening angle of the jet, both of which are uncertain and could change with time. However, if we assume that both of these quantities are constant, the flux we observe and the accretion luminosity of the hole are linearly related. Thus, while the magnitude of the observed flux cannot be determined exactly with our model (without performing a more in-depth analysis of the jet), its qualitative appearance can be reproduced. The value of $M_0$, which is a relevant quantity if the TDE occurs outside the tidal radius, also affects the magnitude of the jet luminosity. The time of disruption is also unknown, as the initial chaotic behavior of the event is not predicted by any model, making the time at which the observed flux reached a maximum incalculable. However, since there is a steady decline after a timescale on the order of days with no recurring rise, it is probable that the maximum fallback rate occurred somewhere near in time to the triggering event.}

{With all of these considerations, we will restrict our attention to times greater than roughly 15 days after the trigger, where the observed flux approximately follows a power law. Given this restriction and our uncertainty in the exact value of the intrinsic flux, the first constraint on our model is that the power-law index of the jet luminosity should be between $1.5$ and $2.2$. The second constraint is that the accretion luminosity produced by the hole should be near the Eddington limit of the hole after about 500 days from the time of the triggering. The luminosity of the jet must also change by about an order of magnitude during this length of time, evident in Figure 1 of \citet{tch13}.}

{If we adopt the model discussed in the beginning of section 4.3, which places the periapsis of the disrupted star exactly at the tidal radius, and we choose the set of fiducial parameters that produced Figure \ref{fig:nplotsmbh}, the only free parameter left is the mass of the black hole. Because the power-law index for later times is around 1.5, there will exist a qualitatively good fit between the model and the data. Our models predict that smaller black holes produce a greater change in the jet luminosity over the duration of the super-Eddington event, and an order-of-magnitude change in the luminosity requires a black hole of mass $M_h \simeq 10^5M_{\astrosun}$. For black holes with masses in this range, the accretion luminosity of the hole is on the order of its Eddington limit around 500 days after the maximum fallback. This prescription is thus broadly consistent with \emph{Swift} J1644+57.}

{If the tidal disruption occurs at a distance such that the star is only partially destroyed, the power-law rate of return is the most appropriate method by which we can analyze the fallback onto the ZEBRA. Since the power-law associated with the accretion luminosity is always between about 1.5 and 1.7, there will exist qualitatively good agreement between the observations of \emph{Swift} J1644+57 and the ZEBRA prediction. For these models, the change in flux being an order of magnitude again requires that $M_h \simeq 10^5M_{\astrosun}$, consistent with the description that places the periapsis of the star at the tidal disruption radius. For these fallback rates, the accretion luminosity is near the Eddington limit of the hole after 500 days from the maximum, though the precise number depends on $M_0$.}

\section{Discussion and conclusions}
We have outlined a novel approach to describing the super-Eddington accretion disks generated during tidal disruption events. Following \citet{loe97}, we used the low specific angular momentum of the tidally-disrupted material to place the material in a nearly-spherically symmetric configuration around the hole. However, instead of forcing a strictly spherical envelope to enclose a thick disk, which we believe to be unstable due to the absorption of energy and transfer of angular momentum, we self-consistently account for the distribution of angular momentum throughout the material. 

In our models, the {accretion energy released by} the black hole and shock heating pump a significant amount of energy into the debris, puffing up the disk. We encounter a point where the Bernoulli parameter approaches zero, leaving a quasi-spherical envelope that is marginally bound. Further energy input would unbind the material, most likely resulting in a wind (see Appendix B). Instead of creating a wind, we posit that the accretion energy of the black hole is instead redirected to the poles, resulting in the {formation} of a jet that serves as the exhaust route for the excess energy. The resulting configuration is a zero-Bernoulli accretion (ZEBRA) flow, threaded by a bipolar jet. This type of object is specifically relevant to the recently observed X-ray transient \emph{Swift} J1644+57.

 {The creation of the jet is a natural consequence of the fact that ZEBRA envelopes are closed up all the way to the poles, leaving no disk surface from which one could launch a wind and remove the super-Eddington accretion luminosity. Because the liberated gravitational energy would have to propagate through the entire system to be released at the photosphere, which is not possible owing to the supercritical nature of its generation rate, it is forced to exit along the poles. Another consequence of the super-Eddington luminosity is the inability of the flow to cool efficiently, forcing it to maintain its zero-Bernoulli nature. The supercritical accretion luminosity thus forms the cornerstone of the consistency of our model.}

Following the analysis of \citet{bla04}, we demonstrated the existence of self-similar solutions to the momentum and Bernoulli equations with $B = 0$, which form a particular subset of the gyrentropic flows discussed by those authors. The gyrentropic nature of our flows is a result only of our assumption of the globally-zero Bernoulli parameter, independent of any stability considerations, such as the presence or absence of MRI.  The ZEBRA flows were shown to close up only exactly at the poles, indicating the quasi-spherical nature of the envelopes. 

We showed that there exists an unspecified parameter, denoted by $q$ (linearly related to the parameter $n$ of \citealt{bla04}), which characterizes the radial gradients of the density and pressure and the sub-Keplerian nature of the flow. For TDEs, the total mass and angular momentum of the progenitor star, coupled to our specification of the trapping radius as the edge of the envelope, determine the value of $q$. For low specific angular momentum, the gradients of the density and pressure increase, approaching the {isentropic} value of a non-rotating star as the angular momentum goes to zero.

ZEBRA envelopes have a radial extent of hundreds to thousands of Schwarzschild radii, validating our neglect of general relativistic effects over the bulk of the flow. However, the excess of angular momentum at small radii to account for the relativistic gravitational field could play a significant role in our determination of the gross properties of the configuration. By using the pseudo-Newtonian potential of \citet{pac80}, we demonstrated that, while the specific angular momentum, pressure, and density can deviate significantly from their Newtonian values in regions close to the hole, the total mass and angular momentum of the envelope are largely unaffected. 

These models apply to the super-Eddington phase of accretion, namely when the flow is unable to cool via radiative losses. We were able to predict the accretion and jet luminosities associated with ZEBRA flows, and found that, indeed, the rates are highly supercritical, providing a self-consistency check on our assumptions. Another aspect of our flows that is asserted a priori is that the Bernoulli parameter is precisely zero, which we know must break down close to the hole. The implications of a non-zero, but constant, $B$ are addressed in Appendix B.  The results derived in the previous sections are shown to be insensitive to this assumption provided that $|{B}| < GM_h/r$.

Because of the very high accretion rates, an appreciable amount of mass is lost on a dynamically relevant timescale. This consideration allowed us to determine the time-dependent nature of the properties of the accretion flow and the jet. {By using an analytic model closely following that of \citet{lod09} to describe the fallback rate of tidally-stripped material}, it was shown that the jet luminosity {roughly follows the rate {at which material returns to pericenter}, but with a few notable differences}. {In addition to using the model for which the pericenter distance of the disrupted star equals the tidal radius, we investigated the consequences of letting the fallback onto the ZEBRA scale as a power-law. This model serves as a proxy for the late evolution of TDEs for which the pericenter of the stellar progenitor lies outside the tidal radius. In these cases, the accretion rate onto the hole also follows a power-law decline, but with a power-law index that is less steep than that of the fallback rate.} We also demonstrated that ZEBRA envelopes produced by TDEs should have {approximately-constant} effective temperatures of $T_{} \simeq 5\times10^{4}$ K, placing the peak of their bolometric luminosities in the far UV to soft X-ray. 

We compared our models with the observed properties of the transient X-ray source \emph{Swift} J1644+57{. Because of the uncertainties in the opening angle of the jet, its Lorentz factor, and the time at which the X-ray flux reached its peak magnitude, we were unable to place many direct constraints on our model. However, we found broad consistency with our models and the observations if the black hole has a mass on the order of $10^5M_{\astrosun}$, assuming a disrupted star of solar type and a constant jet Lorentz factor and opening angle.}}

The existence of a ZEBRA is contingent on the availability of an exhaust route for the excess energy produced in the accretion process. In this account we have presupposed the existence of a jet as this conduit, and we demonstrated its consistency with the source \emph{Swift} J1644+57. We have foregone, however, any explicit analysis concerning its generation or its interaction with the ZEBRA flow. {We also neglected any changes in the Lorentz factor or the beaming angle of the jet, both of which would have observable effects on the X-ray luminosity.} These aspects of the problem will be addressed in a future paper.

In addition to tidal disruption events, ZEBRA flows may manifest themselves in other astrophysical systems. One such application is to failed supernovae, or collapsars \citep{woo93,mac99}.  In this model, a highly evolved, rotating star undergoes a type II supernova. The core collapses directly to a black hole, the remaining stellar material creating an accretion disk and producing a jet. The internal shocks within the jet provide one mechanism capable of producing the gamma rays we observe in long gamma-ray bursts (GRB). Outflows farther from the poles are thought to unbind the envelope and produce the supernova signature observed in many long GRBs \citep{woo06}. However, there have been a few cases in which we observe a GRB devoid of any supernova afterglow \citep{fyn06}, even though the location of the GRB should have provided no impediment (e.g., dust extinction or light contamination) to our observation of the afterglow. It is possible that, in these instances, the outflows away from the poles were not sufficient to unbind the envelope, leaving it intact above the black hole. This environment is precisely that in which a ZEBRA flow would arise, as a wind is unable to be created due to the presence of the overlying stellar material.  As more energy is pumped into the material, the entire mass of the progenitor may come to an approximate equilibrium described by our $B = 0$ prescription. Another application would be in the deep interior of a quasi-star, a giant proto-galactic gas cloud supported by black hole accretion \citep{beg06,beg08}. Because the black hole accretes at the Eddington limit of the total quasi-star, whose mass far exceeds that of the hole, the accretion rate is highly supercritical. The overlying gas prevents the generation of a wind or any other exhaust mechanism, making a ZEBRA flow the most appropriate description of the fluid.

In both of the previous examples, the accretion rates and other physical processes create a natural environment for ZEBRAs. However, they both differ from tidal disruption events in the gravitational role played by the black hole: in TDEs, the black hole dominates the mass of the system, and so the gravitational potential is given by that of a point mass. In a failed supernova, the black hole generated by the collapse is on the order of the mass of the overlying material. Thus, as we move away from the hole into the ZEBRA envelope, there will come a point where the enclosed mass roughly equals that of the hole. The point-mass prescription then becomes invalid. For a quasi-star, the black hole constitutes only a small fraction of the total mass, and hence the self-gravitating nature of the flow must be considered in order to adequately describe the properties of the ZEBRA.

The fluid and Bernoulli equations with an arbitrary gravitational potential may be written down in a straightforward manner, Poisson's equation being the extra constraint that closes the system. An analysis of these relations, in which we self-consistently include both the angular momentum of the gas and its self-gravitating nature, will be deferred to a later paper. 

\acknowledgements
This work was supported in part by NSF grant AST-0907872 and NASA's Fermi Guest Investigator Program. We thank Phil Armitage and Greg Salvesen for comments on an early draft. {The authors would also like to thank the anonymous referee for useful and insightful suggestions.} This work made use of data supplied by the UK Swift Science Data Centre at the University of Leicester.

\appendix

\section{Appendix A: Non-self-similar ZEBRA solutions}
\citet{bla04} demonstrated that, if the angular momentum is distributed in a quasi-Keplerian fashion, i.e. $\ell^2 \propto GMr\sin^2\theta$, then there exist self-similar solutions for the pressure and the density throughout the disk.  Here we wish to demonstrate that the converse of this statement, namely ``If the pressure and density fall off in a self-similar manner, then the angular momentum is quasi-Keplerian," is also true.  In the process we will find the general solution for the density, pressure, and angular momentum distributions of ZEBRA flows in a Keplerian potential.

As a reminder, the momentum and Bernoulli equations governing the ZEBRA flow are

\begin{equation}
\frac{1}{\rho}\frac{\partial{p}}{\partial{r}} = -\frac{GM_h}{r^2}+\frac{\ell^2\csc^2\theta}{r^3}, \label{rmom1}
\end{equation}
\begin{equation}
\frac{1}{\rho}\frac{\partial{p}}{\partial\theta} = \frac{\ell^2\csc^2\theta\cot\theta}{r^2}, \label{thmom1}
\end{equation}
\begin{equation}
-\frac{GM_h}{r}+\frac{\ell^2\csc^2\theta}{2r^2}+\frac{\gamma}{\gamma-1}\frac{p}{\rho} = 0.  \label{bern1}
\end{equation}
Now make the following auxiliary definitions:

\begin{equation}
\frac{\ell^2\csc^2\theta}{r^2} = \frac{GM_h}{r}f(r,\theta),
\end{equation}
\begin{equation}
p(r,\theta) = \frac{GM_h}{r}h(r,\theta). 
\end{equation}
Inserting these definitions into \eqref{rmom1}, \eqref{thmom1}, and \eqref{bern1}, we find the following differential equation for $h$:

\begin{equation}
\frac{1}{\gamma-1}h(r,\theta)+r\frac{\partial{h}}{\partial{r}} = \frac{1}{2}\frac{\partial{h}}{\partial\theta}. 
\end{equation}
The solution to this partial differential equation may be found most simply by separating variables. Doing so, we obtain

\begin{equation}
h(r,\theta) = r^{-\frac{1}{\gamma-1}}\int_0^{\infty}c(\lambda)(r\sin^2\theta)^{\lambda}d\lambda.  \label{heq}
\end{equation}
Here $\lambda$ is the arbitrary constant obtained from the separation of variables technique, and $c(\lambda)$ is the constant of integration which is, in general, a function of $\lambda$.  The total solution, capable of being matched to arbitrary boundary conditions, is then a sum of the eigensolutions appropriate to a single $\lambda$ (if $\lambda$ takes on a discrete set of values, the integral becomes a sum and $c(\lambda) \rightarrow c_{\lambda}$). The range of $\lambda$ has been chosen in hindsight to be consistent with the restriction that the square of the angular momentum and the density both be positive.

Using $\eqref{heq}$, we can readily determine expressions for the density, pressure, and angular momentum, which we find to be

\begin{equation}
\rho(r,\theta) = r^{-\frac{1}{\gamma-1}}\int_0^{\infty}(\lambda+\frac{\gamma}{\gamma-1})c(\lambda)(r\sin^2\theta)^{\lambda}d\lambda, \label{dens1}
\end{equation}
\begin{equation}
p(r,\theta) = GM_hr^{-\frac{\gamma}{\gamma-1}}\int_0^{\infty}c(\lambda)(r\sin^2\theta)^{\lambda}d\lambda,  \label{press1}
\end{equation}
\begin{equation}
\ell^{\,2}(r,\theta) = 2GM_hr\sin^2\theta\frac{\int_0^{\infty}\lambda\,{c(\lambda)}(r\sin^2\theta)^{\lambda}d\lambda}{\int_0^{\infty}(\lambda+\frac{\gamma}{\gamma-1})c(\lambda)(r\sin^2\theta)^{\lambda}d\lambda}.  \label{angmom}
\end{equation}
We are now in a position to prove the statement at the beginning of this appendix: if we require the density or pressure to vary self-similarly, then $c(\lambda) = c'\delta(\lambda-\lambda')$, where $c'$ is a constant independent of $\lambda$ and $\delta(x)$ is the Dirac delta function. Inserting this relation for $c(\lambda)$ into \eqref{angmom}, we find

\begin{equation}
\ell^{\,2}(r,\theta) = \frac{2\lambda}{\lambda+\frac{\gamma}{\gamma-1}}GM_hr\sin^2\theta, 
\end{equation}
which agrees with the result in section 2.2 if we let $\lambda = n-3/2+1/(\gamma-1)$. The angular momentum distribution of a self-similar flow is therefore quasi-Keplerian.  Furthermore, even if $c(\lambda)$ is not a delta function, the functional dependence of the angular momentum is the Keplerian solution multiplied by a ratio of integrals, with each of those integrals having the same leading power of $r$. For this reason the dominant behavior of the angular momentum will always be Keplerian.

\section{Appendix B: Non-zero Bernoulli parameter}
One of the tenets upon which much of our previous analysis rests is that the Bernoulli parameter is exactly zero.  Here we would like to investigate the consequences of letting $B$ become negative, meaning that the disk is more than marginally bound; this situation may occur if the jet turns on before enough energy is pumped into the debris disk.  We will also examine the case where $B > 0$, and we will show that this regime is associated with a wind.

Assuming that we can still regard $B$ as roughly constant, then $\nabla{B} = 0$, and the gyrentropic nature of the flow in the disk is preserved (see Section 2). The fluid and Bernoulli equations are now

\begin{equation}
\frac{1}{\rho}\frac{\partial{p}}{\partial{r}} = -\frac{GM_h}{r^2}+\frac{\ell^2\csc^2\theta}{r^3}, 
\end{equation}
\begin{equation}
\frac{1}{\rho}\frac{\partial{p}}{\partial\theta} = \frac{\ell^2\cot\theta\csc^2\theta}{r^2}, 
\end{equation}
\begin{equation}
-\frac{GM_h}{r}+\frac{\ell^2\csc^2\theta}{2r^2}+\frac{\gamma}{\gamma-1}\frac{p}{\rho} = B,
\end{equation}
where we have written the Bernoulli parameter as $B$, with $B$ a negative number. We can generalize our analysis in the previous appendix to include non-zero $B$. The resultant self-similar solutions for the angular momentum, density, and pressure are

\begin{equation}
\ell^2 = a\bigg{(}\frac{GM_h}{r}+B\bigg{)}r^2\sin^2\theta, 
\end{equation}
\begin{equation}
\rho = \rho_0\bigg{(}\frac{\frac{GM_h}{r}+B}{\frac{GM_h}{r_0}+B}\bigg{)}^{\alpha+\frac{1}{\gamma-1}}\bigg{(}\frac{r^2}{r_0^2}\bigg{)}^{\alpha}(\sin^2\theta)^{\alpha},
\end{equation}
\begin{equation}
p=\beta\rho_0\bigg{(}\frac{GM_h}{r}+B\bigg{)}\bigg{(}\frac{\frac{GM_h}{r}+B}{\frac{GM_h}{r_0}+B}\bigg{)}^{\alpha+\frac{1}{\gamma-1}}\bigg{(}\frac{r^2}{r_0^2}\bigg{)}^{\alpha}(\sin^2\theta)^{\alpha}. 
\end{equation}
Here our notation is consistent with that in section 2, i.e. $a$, $\alpha$, $n$, and $\beta$ retain their original definitions.  We see that a disk with finite binding energy differs {in its radial structure} from that considered previously only in regions where $GM_h/r \sim |B|$, with the angular dependence completely unaltered.  Therefore, our analysis in sections 2-4 concerning the properties of the disk is largely incorrect only if $GM_h/|B| < R$.

To determine when and if this inequality is satisfied, let us assume that the inverse is true, i.e. $R<GM_h/|B|$, so that the results from the preceding sections are almost correct.  Then we can approximate the outer radius by the expression in section 3.  We then find, in order for our neglect of the Bernoulli parameter to be permissible, that $B$ must satisfy

\begin{equation}
|B| < \frac{GM_h}{\bigg{(}\frac{\kappa{y}}{4\pi{c}}\mathscr{M}\sqrt{GM_h}\beta\sqrt{a}(3-q)\bigg{)}^{2/5}}. \label{Beq}
\end{equation}
For our current models, the right-hand side of \eqref{Beq} takes on values that are on the order of $\approx 10^{\,17}$.  To see if this number is consistent with our neglect of finite binding energy, we can further specify the Bernoulli parameter by recalling the gravitational potential energy of the disk and its relation to the star, which implies $B = -\delta(GM_*/R_*)(M_h/M_*)^{1/3}$, where $\delta$ is a numerical factor.  Inserting numbers into \eqref{Beq}, we find that our assumptions in sections 2-4 are correct if $\delta \lesssim 1$.  As we have argued, the shock heating and energy generation in the inner regions of the disk are thought to raise the Bernoulli parameter, so that $\delta \lesssim 1$ should be satisfied in nearly all cases.  However, for larger black hole masses or larger progenitors, the binding energy, and hence the Bernoulli parameter, will increase, and the assumption of $B \approx 0$ may start to break down. 

By changing the sign of $B$, we obtain the solutions for positive-Bernoulli disks. As anticipated, these models yield finite pressure, density, and angular momentum at infinity, confirming our suspicions that positive-Bernoulli disks produce winds. In fact, if $B$ becomes too large, the density again approaches a power law but with $2\alpha$ replacing $-q$. In this limit we can show that, for $q$ that leave the density finite at the poles, all solutions predict an energy that increases as we go out in radius. These two physically meaningless conclusions lead us to the assertion that positive $B$ solutions do not describe wind-less disks, and hence are not appropriate to our modeling of the debris disks produced by tidal disruption events.

\clearpage

\end{document}